\documentclass[
  twocolumn,
  prb,
  amsmath,
  amssymb,
  superscriptaddress,
  floatfix
]{revtex4}

\usepackage{bm}
\usepackage{graphicx}

\begin{document}

\title{Electron transport in disordered graphene}

\author{P.~M.~Ostrovsky}
\affiliation{
 Institut f\"ur Nanotechnologie, Forschungszentrum Karlsruhe,
 76021 Karlsruhe, Germany
}
\affiliation{
 L.~D.~Landau Institute for Theoretical Physics RAS,
 119334 Moscow, Russia
}

\author{I.~V.~Gornyi}
\altaffiliation{
 Also at A.F.~Ioffe Physico-Technical Institute,
 194021 St.~Petersburg, Russia.
}
\affiliation{
 Institut f\"ur Nanotechnologie, Forschungszentrum Karlsruhe,
 76021 Karlsruhe, Germany
}

\author{A.~D.~Mirlin}
\altaffiliation{
 Also at Petersburg Nuclear Physics Institute,
 188350 St.~Petersburg, Russia.
}
\affiliation{
 Institut f\"ur Nanotechnologie, Forschungszentrum Karlsruhe,
 76021 Karlsruhe, Germany
}
\affiliation{
 Institut f\"ur Theorie der kondensierten Materie,
 Universit\"at Karlsruhe, 76128 Karlsruhe, Germany
}

\date{\today}

\begin{abstract}
We study electron transport properties of a monoatomic graphite layer
(graphene) with different types of disorder. We show that the transport
properties of the system depend strongly on the character of disorder. Away
from the half filling, the concentration dependence of conductivity is linear
in the case of strong scatterers, in line with recent experimental
observations, and logarithmic for weak scatterers. At half filling the
conductivity is of the order of $e^2/h$ if the randomness preserves one of the
chiral symmetries of the clean Hamiltonian; otherwise, the conductivity is
strongly affected by localization effects.
\end{abstract}

\maketitle

\section{Introduction}
\label{s1}

Recently, Novoselov \textit{et al} have succeeded in fabrication of monolayer
graphite (graphene) samples.\cite{Novoselov04} Subsequent transport
measurements\cite{Novoselov05, Zhang, Novoselov06, Morozov06, Zhang06} have
shown that graphene is a conductor with remarkable electronic properties. These
experimental discoveries have triggered an outbreak of theoretical activity;
see, in particular, Refs.\ \onlinecite{GusyninCo, Kane05, Katsnelson, KNG,
Guinea, Morpurgo06, brey, Abanin06, Pogorelov06, Beenakker06, Tworzydlo06,
Titov06, McCannFalko, McCannPRL06, Cheianov06, Koshino06, Ando06,
Khveshchenko06loc, Foster06, McCannCondMat06, NomuraFerro, Nomura06,
Falkovsky06, Cheianov06Friedel, AleinerEfetov, Altland06, McCann-alone,
Sinitsyn, Lukyanchuk, Snyman, Dahal, Wehling}. Charge carriers in graphene have
a relativistic (Dirac) spectrum,\cite{Wallace, Ando05Review} which makes the
transport properties of this material highly interesting from the point of view
of both fundamental physics and potential applications. It is widely believed
that graphene-based devices may be of outstanding importance for future
nanoelectronics.

This work has been motivated by the following two experimental
observations.\cite{Novoselov05, Zhang} First, it was found that the graphene
conductivity is linear in the concentration of carriers (counted from the half
filling) with a high accuracy. Second, it was found that at half filling the
conductivity (per spin direction and per valley) is close to $e^2/h$ and does
not show any definite temperature dependence in a broad temperature range.  The
aim of this paper is to analyze what one should expect for conductivity from
the theoretical point of view and whether these theoretical predictions may be
compatible with experimental findings. We will see that, in view of the
unconventional character of the graphene spectrum, the theoretical results
depend crucially on the nature of disorder.

The structure of the paper is as follows. In Sec.\ \ref{s2} we introduce the
model describing electronic properties of graphene with various types of
disorder. In Sec.\ \ref{s3} we analyze the dependence of conductivity on the
electron concentration away from the half-filling point. We consider the two
limits of weak and strong scatterers and construct the corresponding ``phase
diagram''. Section \ref{s4} is devoted to the conductivity at half filling
under the assumption that the disorder preserves one of the chiral symmetries
of the Dirac Hamiltonian. Our findings are summarized in Sec.\ \ref{s5}. Some
technical details are presented in two Appendices.

\section{The Model}
\label{s2}

\subsection{Clean graphene}
\label{s2.1}

\begin{figure}
\includegraphics[width=\columnwidth]{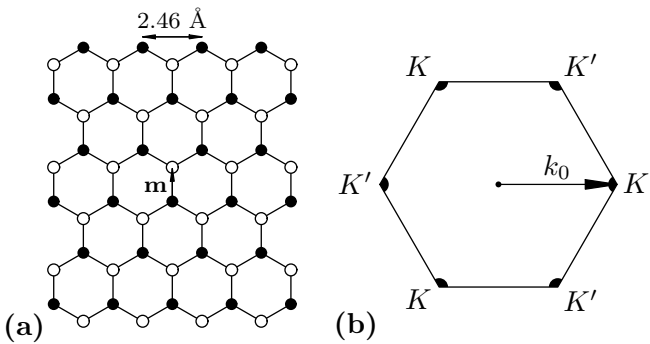}
\caption{(a) Honeycomb lattice of carbon atoms of graphene. Solid and open
circles denote the atoms of $A$ and $B$ sublattices respectively. (b) The first
Brillouin zone of graphene. The nodal points of the spectrum are located in the
corners of the zone. The two nonequivalent nodal points are denoted as $K$ and
$K'$.}
\label{Fig:honey}
\end{figure}

The carbon atoms of graphene are arranged in the honeycomb lattice (see Fig.\
\ref{Fig:honey}a) with the period $a = 2.46$~\AA. Each carbon atom of intrinsic
graphene has one valence electron forming the $\pi$-bonds to the three
neighbors. The electronic spectrum of graphene is well described by the
tight-binding model\cite{Wallace} taking into account the nearest-neighbor
hopping. The first Brillouin zone for this system has a form of a hexagon (see
Fig.\ \ref{Fig:honey}b) with the distance $k_0 = 2h/3a$ from the center to the
apex. The honeycomb lattice contains two sites per elementary cell. This
permits the grouping of all the atoms into two sublattices, $A$ and $B$. The
nearest neighbors of an atom from the sublattice $A$ belong to the sublattice
$B$ and vice versa. The symmetry group of the honeycomb lattice contains an
element swapping the two sublattices. Hence, for each value of quasimomentum
$\mathbf k$ within the Brillouin zone, two states exist with the energies $\pm
E(\mathbf k)$. These two spectrum branches are degenerate at the isolated
points in the corners of the Brillouin zone, $E(k_0) = 0$.  With one electron
per site the system is exactly in the half-filling state when the nodal points
of the spectrum lie at the Fermi level. Among six apices of the hexagonal
Brillouin zone only two are nonequivalent. They are referred to as $K$ and
$K'$. The electrons with momentum close to these two points, and hence with low
energy, are relevant in studying the physics of the system for electron
concentrations not too far from the half filling.

The tight-binding Hamiltonian is a $4 \times 4$ matrix operating in the AB
space of the two sublattices and in the $K$--$K'$ space of the valleys.
Therefore we introduce the four-component wave function
\begin{equation}
 \Psi
  = \{\phi_{AK}, \phi_{BK}, \phi_{BK'}, \phi_{AK'}\}^T.
 \label{4D}
\end{equation}
In this representation the Hamiltonian has the form
\begin{equation}
 H
  = v_0 \tau_3 \bm{\sigma}\mathbf{k}.
 \label{ham}
\end{equation}
Here $\tau_3$ is the third Pauli matrix in the $K$--$K'$ space and $\bm{\sigma}
= \{\sigma_1, \sigma_2\}$ is the two-dimensional vector of Pauli matrices in
the AB space. The Fermi velocity in graphene is $v_0 \simeq 10^8$ cm/s. In
fact, the form of the Hamiltonian (\ref{ham}) is universal and does not rely on
the tight-binding approximation. The degeneracy of the spectrum in $K$ and $K'$
points is provided by the two-dimensional representation of the honeycomb
lattice symmetry group while the expression (\ref{ham}) is the first-order
$k$-expansion near these points. As $k$ is increased the higher-order
nonuniversal terms of this expansion come into play. For our purposes, it will
be sufficient to introduce the high energy cut-off $\Delta$ and to assume the
spectrum to be linear up to $|\mathbf{k}| = \Delta/v_0$. Indeed, all divergent
momentum integrals appearing below have the logarithmic character; thus,
details of the high-energy regularization are irrelevant. The Green function
for the Hamiltonian (\ref{ham}) of the clean graphene reads
\begin{equation}
 G_0^{R(A)}(\varepsilon, \mathbf{k})
  = \frac{\varepsilon + v_0 \tau_3 \bm{\sigma}\mathbf{k}}{(\varepsilon
    \pm i0)^2 - v_0^2 k^2}.
 \label{Green0}
\end{equation}

\subsection{Potential disorder}
\label{s2.2}

We incorporate now disorder in the model. Let us consider first the impurities
modifying the potential on nearby lattice sites. A detailed description due to
McCann and Fal'ko\cite{McCannFalko} contains 10 real parameters for the
potential of a single impurity. In the present paper we will use the simplified
model introduced by Shon and Ando in Ref.\ \onlinecite{ShonAndo}, which retains
the essential physics of the problem. This model treats impurities in the
framework of the same tight-binding approximation as was used for the pure
system. An impurity is placed at a site of the lattice and has a potential
$U(\mathbf{r})$. We use the two discrete Fourier transforms of this function 
with respect to the two sublattices.
\begin{align}
 U_\mathbf{q}
  &= \frac{\sqrt 3 a^2}{2} \sum_{\mathbf{r}}
     U(\mathbf{r})\, e^{-i \mathbf{q} \mathbf{r}}, \\
 U'_\mathbf{q}
  &= \frac{\sqrt 3 a^2}{2} \sum_{\mathbf{r}}
     U(\mathbf{r}-\mathbf{m})\, e^{-i \mathbf{q} \mathbf{r}}.
\end{align}
The summation runs over all elementary cells of the honeycomb lattice, and
the vector $\mathbf{m}$ points from the $A$ sublattice site to the $B$
sublattice site of the same elementary cell (see Fig.\ \ref{Fig:honey}a). The
quantity $U_\mathbf{q}$ is the scattering amplitude for the electrons of the
same sublattice where the impurity resides, while $U'_\mathbf{q}$ is the
scattering amplitude for the electrons of the other sublattice.

Assuming $U_\mathbf{q}$ and $U'_\mathbf{q}$ are slow functions of the momentum
$\mathbf{q}$, we keep only two values of these amplitudes for intravalley,
$U_0$ and $U'_0$, and intervalley, $U_{k_0}$ and $U'_{k_0}$, scattering. The
hexagonal symmetry of the honeycomb lattice makes the amplitude $U'_{k_0}$
vanish while the three other amplitudes are real. Thus we are left with the
three parameters of an impurity potential. This is straightforward to put them
in a matrix in the $4$-dimensional representation (\ref{4D}). If the impurity
site belongs to the sublattice A and to the elementary cell $\mathbf{r}_i$, the
scattering matrix takes the form
\begin{equation}
\label{VAri}
 V^{\mathrm{A}}_{\mathbf{q}}(\mathbf{r}_i)
  = \begin{pmatrix}
      U_0 & 0 & 0 & U_{k_0} e^{-2i \mathbf{k}_0 \mathbf{r}_i} \\
      0 & U'_0 & 0 & 0 \\
      0 & 0 & U'_0 & 0 \\
      U_{k_0} e^{2i \mathbf{k}_0 \mathbf{r}_i} & 0 & 0 & U_0 \\
    \end{pmatrix} e^{-i \mathbf{q} \mathbf{r}_i}.
\end{equation}
For the impurity located in the sublattice B, we have
\begin{equation}
\label{VBri}
 V^{\mathrm{B}}_{\mathbf{q}}(\mathbf{r}_i)
  = \begin{pmatrix}
      U'_0 & 0 & 0 & 0 \\
      0 & U_0 & U_{k_0} e^{-2i \mathbf{k}_0 \mathbf{r}_i} & 0 \\
      0 & U_{k_0} e^{2i \mathbf{k}_0 \mathbf{r}_i} & U_0 & 0 \\
      0 & 0 & 0 & U'_0 \\
    \end{pmatrix} e^{-i \mathbf{q} \mathbf{r}_i}.
\end{equation}

If the potential disorder is weak and obeys Gaussian distribution, the only
relevant quantity is the auto-correlation function of the second order $\left<
V_\mathbf{q} \otimes V_{-\mathbf{q}} \right>$. We denote the impurity
concentration by $n_\text{imp}$ and obtain after averaging with respect to
positions of the impurities
\begin{multline}
 \left< V_\mathbf{q} \otimes V_{-\mathbf{q}} \right> \\
  = \frac{n_\text{imp}}{2} \left<
      V^{\mathrm{A}}_{\mathbf{q}}(\mathbf{r}_i)
      \otimes V^{\mathrm{A}}_{-\mathbf{q}}(\mathbf{r}_i)
      + V^{\mathrm{B}}_{\mathbf{q}}(\mathbf{r}_i)
      \otimes V^{\mathrm{B}}_{-\mathbf{q}}(\mathbf{r}_i)
    \right> \\
  = 2 \pi v_0^2 \Bigl\{
      \alpha_0\; \sigma_0 \tau_0 \otimes \sigma_0 \tau_0
      + \gamma_z\; \sigma_3 \tau_3 \otimes \sigma_3 \tau_3 \\
      + \frac{\beta_\perp}{4} \bigl[
        \sigma_1 \tau_1 \otimes \sigma_1 \tau_1 
        + \sigma_1 \tau_2 \otimes \sigma_1 \tau_2 \\
        + \sigma_2 \tau_1 \otimes \sigma_2 \tau_1
        + \sigma_2 \tau_2 \otimes \sigma_2 \tau_2
      \bigr]
    \Bigr\}.
\end{multline}
Here we introduce the following three dimensionless parameters
\begin{subequations}
\begin{align}
 \alpha_0
  &= \frac{n_\text{imp}}{8 \pi v_0^2} (U_0 + U'_0)^2, \\
 \gamma_z
  &= \frac{n_\text{imp}}{8 \pi v_0^2} (U_0 - U'_0)^2, \\
 \beta_\perp
  &= \frac{n_\text{imp}}{4 \pi v_0^2} U_{k_0}^2.
\end{align}
\label{pot_disord}
\end{subequations}
While the notations in Eq.\ (\ref{pot_disord}) may seem strange at this stage,
they will be explained later when we consider randomness of a broader class. To
further simplify the calculations, we will concentrate on the two limiting
cases of short- and long-range potential disorder.\cite{ShonAndo}

The \emph{short-range} impurity scatters electrons in the same sublattice only.
It is equivalent to a potential shift at a particular lattice site. The
amplitudes are $U_0 = U_{k_0} = U/2$ and $U'_0 = 0$. Thus we are left with a
single parameter $U$. The parameters Eqs.\ (\ref{pot_disord}) obey the relation
$\alpha_0 = \gamma_z = \beta_\perp/2$. 

The \emph{long-range} impurity scatters electrons in both sublattices equally
but only within one valley. The scattering length is large in comparison with
the lattice constant but is still smaller than the Fermi wavelength. The
amplitudes are $U_0 = U'_0 = U$, $U_{k_0} = 0$. We have again the single
parameter $U$ as in the case of short-range disorder. Among the parameters
(\ref{pot_disord}) only $\alpha_0$ is not zero in this case.

\subsection{Generic disorder and chiral symmetries}
\label{s2.3}

Let us turn now to the analysis of the symmetries of the clean graphene
Hamiltonian (\ref{ham}). First, the system is obviously uniform and isotropic.
Any disorder considered in this paper preserves these symmetries on average, so
we do not pay much attention to them here. Second, due to the two valley
structure of the electron spectrum the whole SU(2) symmetry group exists in
an \emph{isospin} space of the valleys. The generators of these group
are\cite{McCannCondMat06}
\begin{equation}
 \Lambda_x
  = \sigma_3 \tau_1,
 \qquad
 \Lambda_y
  = \sigma_3 \tau_2,
 \qquad
 \Lambda_z
  = \sigma_0 \tau_3.
 \label{Lambda-matrices}
\end{equation}
These three operators commute with the Hamiltonian and anticommute with each
other. There are other three matrices $\Sigma_{x,y,z}$ introduced in Ref.
\onlinecite{McCannCondMat06}
\begin{equation}
 \Sigma_x
  = \sigma_1 \tau_3,
 \qquad
 \Sigma_y
  = \sigma_2 \tau_3,
 \qquad
 \Sigma_z
  = \sigma_3 \tau_0.
 \label{Sigma-matrices}
\end{equation}
These operators generate an additional SU(2) group of a \emph{pseudospin}.
They \emph{do not} commute with the Hamiltonian (\ref{ham}), however any of
these matrices commute with any of $\Lambda_{x,y,z}$. 

Third, the time inversion operation (we denote it $T_0$) in the
representation Eq.\ (\ref{4D}) reads
\begin{subequations}
\begin{equation}
 T_0:\quad A
  \mapsto \sigma_1 \tau_1 A^T \sigma_1 \tau_1.
\end{equation}
The Hamiltonian Eq.\ (\ref{ham}) is invariant under time inversion (note that
the momentum operator changes sign under transposition). Combining the $T_0$
operation with any of $\Lambda_{x,y,z}$ from Eq.\ (\ref{Lambda-matrices}) we
produce three additional symmetry operations
\begin{align}
 T_x:&\quad A
  \mapsto \sigma_2 \tau_0 A^T \sigma_2 \tau_0, \\
 T_y:&\quad A
  \mapsto \sigma_2 \tau_3 A^T \sigma_2 \tau_3, \\
 T_z:&\quad A
  \mapsto \sigma_1 \tau_2 A^T \sigma_1 \tau_2.
\end{align}
\end{subequations}

Finally, there is one more --- namely, \emph{chiral} --- symmetry $C_0$, 
and its three counterparts generated by simultaneous application of $C_0$ and
$\Lambda_{x,y,z}$:
\begin{subequations}
\begin{align}
 C_0:&\quad A
  \mapsto -\sigma_3 \tau_0 A \sigma_3 \tau_0, \\
 C_x:&\quad A
  \mapsto -\sigma_0 \tau_1 A \sigma_0 \tau_1, \\
 C_y:&\quad A
  \mapsto -\sigma_0 \tau_2 A \sigma_0 \tau_2, \\
 C_z:&\quad A
  \mapsto -\sigma_3 \tau_3 A \sigma_3 \tau_3.
\end{align}
\label{chiral}
\end{subequations}
The chiral symmetry $C_0$ can be viewed as the basic chiral symmetry 
of the Hamiltonian (\ref{ham}). Indeed, $C_0$ is distinguished by the 
fact that it is directly produced by the Hamiltonian (\ref{ham}) as
$i\sigma_3 \tau_0 = v_0^{-2}(\partial H/\partial k_x)(\partial H/\partial
k_y)$, while other chiral symmetries require a rotation in the isospin space. 

Generally the chiral symmetry implies that the Hamiltonian takes
block-off-diagonal form under a proper unitary transformation. A generic
disorder preserving $C_z$ symmetry can have only \emph{off-diagonal} matrix
elements in the AB space of sublattices. Some specific examples of chiral
symmetry are (i) bond disorder due to distortions of the lattice ($C_z$
symmetry), (ii) random magnetic field (all four symmetries $C_{0,x,y,z}$),
(iii) dislocations, that are equivalent to a random non-Abelian gauge
field\cite{Morpurgo06, OldGuinea} ($C_0$),
(iv) infinitely strong short-range on-site impurities ($C_z$). In the latter
case an electron cannot occupy the impurity site, implying that all the bonds
adjacent to the impurity are effectively cut. Any potential disorder other than
the described extreme case violates all chiral symmetries. The symmetry is also
broken by a non-zero chemical potential. Thus, the impact of the chiral
character of disorder will be particularly important at the degeneracy point
$\varepsilon = 0$. In Sec.\ \ref{s4} we consider various effects of chiral
symmetry on the density of states and conductivity of graphene.

The average isotropy of the disordered graphene implies that $\Lambda_x$ and
$\Lambda_y$ symmetries of the Hamiltonian are present or absent simultaneously.
Below we combine them into a single notation $\Lambda_\perp$, and proceed in
the same way with $T_\perp$ and $C_\perp$. In Table \ref{Tab:sym} we list all
possible matrix structures of the disorder [in the representation defined
by Eq.\ (\ref{4D})] along with their symmetries. There are 9 different
structures altogether.\cite{McCannCondMat06} Those 5 of them that do not
violate time inversion symmetry coincide\cite{footnoteAE} with ones
considered by Aleiner and Efetov.\cite{AleinerEfetov} We also give the
notations of Ref.\ \onlinecite{Guruswamy}, where the disordered Dirac
Hamiltonian obeying $C_z$ chiral symmetry was considered.

\begin{table*}

\caption{The symmetries of various disorders in graphene. The first five rows
of the table contain disorders preserving time inversion symmetry. They were
considered in Ref.\ \protect\onlinecite{AleinerEfetov}. Next four rows are
occupied by disorders violating time inversion symmetry. We present the matrix
structure of the disorder in two forms: by matrices $\sigma_i \tau_j$ and by
matrices $\Sigma_i \Lambda_j$ as in Ref.\ \protect\onlinecite{McCannCondMat06}.
The notations we use for the amplitudes of the disorder in Gaussian limit are
listed in the third column. The letters $\alpha$, $\beta$, and $\gamma$
correspond to $\Lambda_0$, $\Lambda_{x,y}$, and $\Lambda_z$ components of
the disorder Hamiltonian respectively, while the subscripts $0$, $\perp$, and
$z$ indicate the structure in $\Sigma_0$, $\Sigma_{x,y}$, and $\Sigma_z$
domain. In the fourth and fifth column we give alternative notations from
Refs.\ \protect\onlinecite{AleinerEfetov} and \protect\onlinecite{Guruswamy}.
Our notations are close to those of Ref.\ \protect\onlinecite{AleinerEfetov};
the only difference is in the case of a fully diagonal potential: our parameter
$\alpha_0$ corresponds to $\gamma_0$ from Ref.\
\protect\onlinecite{AleinerEfetov}, while we use $\gamma_0$ for the disorder
$\sigma_0 \tau_3$ discriminating the two valleys.}
\label{Tab:sym}

\begin{ruledtabular}
\begin{tabular}{cclccclcclccclccc}
 \multicolumn{2}{c}{Disorder structure} &&
 \multicolumn{3}{c}{Disorder strength} &&
 \multicolumn{10}{c}{Hamiltonian symmetries}
\\
 $\sigma_i \tau_j$ &
 $\Sigma_i \Lambda_j$ &&
 This paper &
 Ref.\ \onlinecite{AleinerEfetov} &
 Ref.\ \onlinecite{Guruswamy} &&
 $\Lambda_\perp$ & $\Lambda_z$ &&
 $T_0$ & $T_\perp$ & $T_z$ &&
 $C_0$ & $C_\perp$ & $C_z$
\\ \hline \hline
 $\sigma_0 \tau_0$ & $\Sigma_0 \Lambda_0$ &&
 $\alpha_0$ & $\gamma_0/2\pi
v^2$ & &&
 $+$ & $+$ &&
 $+$ & $+$ & $+$ &&
 $-$ & $-$ & $-$
\\
 $\sigma_{\{1,2\}} \tau_{\{1,2\}}$ & $\Sigma_{\{x,y\}} \Lambda_{\{x,y\}}$ &&
 $\beta_\perp$ & $2\beta_\perp/\pi v^2$ & &&
 $-$ & $-$ &&
 $+$ & $-$ & $-$ &&
 $+$ & $-$ & $-$
\\
 $\sigma_{1,2} \tau_0$ & $\Sigma_{x,y} \Lambda_z$ &&
 $\gamma_\perp$ & $\gamma_\perp/\pi v^2$ & $g_A$ &&
 $-$ & $+$ &&
 $+$ & $-$ & $+$ &&
 $+$ & $-$ & $+$
\\
 $\sigma_0 \tau_{1,2}$ & $\Sigma_z \Lambda_{x,y}$ &&
 $\beta_z$ & $\beta_z/\pi v^2$ & $\sqrt{2} g_m$ &&
 $-$ & $-$ &&
 $+$ & $-$ & $-$ &&
 $-$ & $-$ & $+$
\\
 $\sigma_3 \tau_3$ & $\Sigma_z \Lambda_z$ &&
 $\gamma_z$ & $\gamma_z/2\pi v^2$ & &&
 $-$ & $+$ &&
 $+$ & $-$ & $+$ &&
 $-$ & $+$ & $-$
\\ \hline 
 $\sigma_3 \tau_{1,2}$ & $\Sigma_0 \Lambda_{x,y}$ &&
 $\beta_0$ & & $\sqrt{2} g_\mu$ &&
 $-$ & $-$ &&
 $-$ & $-$ & $+$ &&
 $-$ & $-$ & $+$
\\
 $\sigma_0 \tau_3$ & $\Sigma_0 \Lambda_z$ &&
 $\gamma_0$ & & &&
 $-$ & $+$ &&
 $-$ & $+$ & $-$ &&
 $-$ & $+$ & $-$
\\
 $\sigma_{1,2} \tau_3$ & $\Sigma_{x,y} \Lambda_0$ &&
 $\alpha_\perp$ & & $g_{A'}$ &&
 $+$ & $+$ &&
 $-$ & $-$ & $-$ &&
 $+$ & $+$ & $+$
\\
 $\sigma_3 \tau_0$ & $\Sigma_3 \Lambda_0$ &&
 $\alpha_z$ & & &&
 $+$ & $+$ &&
 $-$ & $-$ & $-$ &&
 $-$ & $-$ & $-$
\end{tabular}
\end{ruledtabular}

\end{table*}

\section{Conductivity far from the degeneracy point}
\label{s3}

In this section, we will study the concentration dependence of the conductivity
far from half filling, when the size of Fermi circles around $K$ and $K'$
points is large in comparison with the inverse mean free path. The
dimensionless Drude conductivity (measured in units of $e^2/h$) is then
large,\cite{ShonAndo} so that, at realistic temperatures, one can neglect as a
first approximation the quantum corrections related to localization.  As a
starting point, we will employ the \emph{self-consistent $T$-matrix
approximation} (SCTMA\cite{SCTMARefs}) that takes into account all orders of
scattering at an impurity. It will allow us to study the whole ``phase
diagram'' including the limits of weak (Born) and strong (unitary) scatterers
and the crossover between them. We will discuss the status of SCTMA in Sec.\
\ref{s3.1}, where we will show that while it is not quantitatively justified in
the Born regime, it yields a qualitatively correct behavior of the
conductivity far from the degeneracy point.

\subsection{Self-consistent $T$-matrix approximation}
\label{s3.0}

\subsubsection{Potential disorder}
\label{s3.0.1}

\begin{figure}
\centerline{\includegraphics{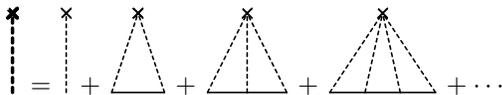}}
\caption{Graphical representation of the $T$ matrix describing the electron
scattering off an impurity.}
\label{Fig:Tmatrix}
\end{figure}

Let us first consider the disorder induced by randomly located
impurities which create the potential given by Eqs.\ (\ref{VAri}) and
(\ref{VBri}).
The sum of all scattering orders determines the complete impurity's $T$ matrix,
as represented graphically in Fig.\ \ref{Fig:Tmatrix}. Averaging the $T$ matrix
with respect to the position of the impurity, we find
\begin{multline}
 \langle T(\varepsilon) \rangle
  = \frac{1}{4} \Biggl[
      \frac{2 U'_0}{1 - U'_0 g} \\
      +\frac{U_0 + U_{k_0}}{1 - (U_0 + U_{k_0}) g}
      +\frac{U_0 - U_{k_0}}{1 - (U_0 - U_{k_0}) g}
    \Biggr]
 \label{T}
\end{multline}
with $g$ being the integral of the Green function,
\begin{equation}
 g(\varepsilon)
  = \int \frac{d^2 k}{(2\pi)^2} G(\varepsilon, \mathbf{k}).
 \label{gint}
\end{equation}
This quantity has trivial matrix structure due to the angular integration. The
electron's self energy is determined by the average value of the $T$-matrix Eq.\
(\ref{T})
\begin{equation}
 \Sigma(\varepsilon)
  =  n_\text{imp} \langle T(\varepsilon) \rangle,
       \label{SigmaT}
\end{equation}
where $n_\text{imp}$ is the concentration of impurities.

Inserting (\ref{SigmaT}) into the bare Green function (\ref{Green0}),
\begin{equation}
 G(\varepsilon, \mathbf{k})
  = \frac{\varepsilon - \Sigma(\varepsilon) + v_0 \tau_3 \bm{\sigma}\mathbf{k}}
    {\bigl[\varepsilon - \Sigma(\varepsilon)\bigr]^2 - v_0^2 k^2},
\end{equation}
and calculating the momentum integral in Eq.\ (\ref{gint}), we get
\begin{equation}
 g(\varepsilon)
  = -\frac{\varepsilon - \Sigma(\varepsilon)}{4 \pi v_0^2}
    \log \frac{- \Delta^2}{\bigl[\varepsilon - \Sigma(\varepsilon)\bigr]^2}.
 \label{g}
\end{equation}
The logarithmic divergence is cut at the momentum $\Delta/v_0$. The sign of the
imaginary part of $g(\varepsilon)$, and hence of the self energy, is determined
by the type of the Green function (advanced or retarded) we are considering.

The equations (\ref{T}), (\ref{SigmaT}), and (\ref{g}) form a closed set that
self-consistently determines the self energy $\Sigma(\varepsilon)$. These
equations take into account all the diagrams with non-intersecting impurity
lines. In the two extreme cases of short- and long-range potential impurities
the self-consistency equation reduces to the form
\begin{equation}
 \Sigma(\varepsilon)
  = \begin{cases}
      \dfrac{n_\text{imp} U}{1 - U g(\varepsilon)},
        & \text{long-range}; \\[10pt]
      \dfrac{n_\text{imp} U}{4 \bigl[ 1 - U g(\varepsilon) \bigr]},
        & \text{short-range}.
    \end{cases}
 \label{SCTMA}
\end{equation}

Once these equations are solved, one can find the density of states and the
conductivity of the system. The density of states (per one spin component) is
\begin{equation}
 \rho(\varepsilon)
  = -\frac{1}{\pi} \mathop{\mathrm{Im}} \mathop{\mathrm{Tr}} \int
    \frac{d^2 k}{(2 \pi)^2}\, G^R(\varepsilon, \mathbf{k})
  = -\frac{4}{\pi} \mathop{\mathrm{Im}} g^R(\varepsilon).
 \label{DoS}
\end{equation}

The conductivity at zero frequency and wave vector is given by the Kubo
formula,
\begin{multline}
\sigma^{\alpha\beta}(\varepsilon)
  = \frac{2}{\pi} \int d^2(r-r') \\
    \times \mathop{\mathrm{Tr}} \Bigl<
      j^\alpha \mathop{\mathrm{Im}} G^R(\varepsilon; \mathbf{r},\mathbf{r'})
      j^\beta \mathop{\mathrm{Im}} G^R(\varepsilon; \mathbf{r'},\mathbf{r})
    \Bigr>.
 \label{Kubo}
\end{multline}
Due to the linear dependence of the Hamiltonian on $\mathbf{k}$, the current
operator is independent of the momentum:
\begin{equation}
 \mathbf{j}
  = e \frac{\partial H}{\partial \mathbf{k}}
  = e v_0 \tau_3 \bm{\sigma}.
 \label{current}
\end{equation}
This results in the absence of the diamagnetic term in the expression for
conductivity.

Equation (\ref{Kubo}) includes averages of the type $\langle j G^R j G^A
\rangle$ along with $\langle j G^R j G^R\rangle$ and $\langle j G^A j G^A
\rangle$. The first one is large in the metallic regime when the energy
$\varepsilon$ is far from the degeneracy point, while the two others give a
contribution of the order of conductance quantum $e^2/h$. Therefore we will
neglect those two in this section.

\begin{figure}
\centerline{\includegraphics{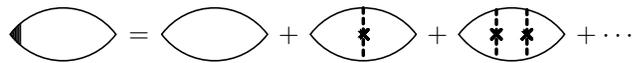}}
\caption{Diagrams for the Drude conductivity including the vertex correction.}
\label{Fig:Drude}
\end{figure}

As discussed above, we will use the Drude approximation for the conductivity,
neglecting weak localization corrections. Graphically, this is equivalent to
the summation of the diagrams shown in Fig.\ \ref{Fig:Drude}. Due to the vector
nature of the current operator in the vertex, only the diagonal parts of the
Green functions contribute to the result. We introduce the notation
\begin{equation}
 \Pi^{RA}(\varepsilon)
  = \int \frac{d^2 k}{(2\pi)^2}
    \mathop{\mathrm{diag}} G^R(\varepsilon, \mathbf{k})
    \mathop{\mathrm{diag}} G^A(\varepsilon, \mathbf{k}).
 \label{Pi}
\end{equation}
The sum of the ladder diagrams in Fig.\ \ref{Fig:Drude} give the correction to
the current vertex. We will use a special notation $\mathcal{V}$ for this
vertex correction factor. In the limit of the short-range potential disorder,
we have $\mathcal{V} = 1$. In the opposite long-range case, the summation of
ladder diagrams yields
\begin{equation}
 \mathcal{V}
  = \frac{1}{1 - \frac{n_\text{imp} U^2}{|1 - U g|^2} \Pi^{RA}}.
 \label{vertex_correct}
\end{equation}
The resulting Drude conductivity has the form
\begin{equation}
 \sigma(\varepsilon)
  = \frac{4}{\pi} e^2 v_0^2 \mathcal{V} \Pi^{RA}.
 \label{Drude}
\end{equation}
In the following sections we will solve the self-consistency equations and find
the density of states and the conductivity in various limits.
We will also analyze the corrections to the SCTMA coming from 
diagrams with intersecting disorder lines.

\subsubsection{Generic Gaussian disorder}
\label{s3.0.2}

In the case of generic weak disorder, one can use a more general equation for
the self energy taking into account all possible disorder amplitudes (listed in
Table \ref{Tab:sym}) in the framework of the Born approximation,
\begin{equation}
 \Sigma(\varepsilon)
  = 2 \pi v_0^2 \alpha\; g(\varepsilon).
 \label{SigmaTGauss}
\end{equation}
Here $\alpha$ is the total strength of the disorder, i.e. the sum of all
amplitudes from Table \ref{Tab:sym},
\begin{equation}
 \alpha
  = \alpha_0 + \beta_0 + \gamma_0
    + \alpha_\perp + \beta_\perp + \gamma_\perp
    + \alpha_z + \beta_z + \gamma_z.
 \label{alpha}
\end{equation}
This quantity is relevant for thermodynamic properties of the system with
Gaussian disorder. The vertex correction is given by
\begin{equation}
 \mathcal{V}
  =  \dfrac{1}{1 - 4 \pi v_0^2 (\alpha - \alpha_\text{tr}) \Pi^{RA}},
        \label{vertex_correctGauss}
\end{equation}
where  
\begin{equation}
 \alpha_\text{tr}
  = \frac{1}{2} \bigl( \alpha_0 + \beta_0 + \gamma_0 \bigr)
    + \alpha_\perp + \beta_\perp + \gamma_\perp
    + \frac{3}{2} \bigl( \alpha_z + \beta_z + \gamma_z \bigr).
 \label{alpha_tr}
\end{equation}
As will be seen below [Eq.\ (\ref{DrudeSCBA})],
$\alpha_\text{tr}$ governs the transport properties of the
Gaussian disordered system.

Using Eqs.\ (\ref{pot_disord}), (\ref{alpha}), and  (\ref{alpha_tr}),
we get for weak long-range and short-range potential disorder considered
above in Sec.\ \ref{s3.0.1}
\begin{align}
 \alpha
  &= 2\alpha_\text{tr}
   = \frac{n_\text{imp}\, U^2}{2 \pi v_0^2}, 
       && \text{long-range};
 \label{alpha-long-range} \\
 \alpha
  &= \alpha_\text{tr}
   = \frac{n_\text{imp}\, U^2}{8\pi v_0^2},
       && \text{short-range}.
\end{align}

\subsection{Born limit: Weak scatterers}
\label{s3.1}

\subsubsection{Self-consistent Born approximation}
\label{s3.1.1}

The simplest situation is the Born limit of weak scattering. Only the lowest
scattering order is relevant in this case. We expand the right-hand side of the
equation (\ref{SCTMA}) to the second order in the scattering amplitude $U$.
(The first-order term is a constant which is absorbed in renormalization of the
chemical potential.) This yields the \emph{self-consistent Born approximation}
(SCBA).\cite{ShonAndo, Ando98, SCBARefs} In this limit, we deal with a generic
weak disorder described by all 9 parameters listed in Table \ref{Tab:sym}. The
results for the special case of potential disorder are easily restored from the
general results with the help of Eqs.\ (\ref{pot_disord}).

The SCBA equation has the form
\begin{equation}
 \Sigma(\varepsilon)
  = -\frac{\alpha}{2} \bigl[\varepsilon - \Sigma(\varepsilon)\bigr]
    \log \frac{- \Delta^2}{\bigl[\varepsilon - \Sigma(\varepsilon)\bigr]^2}.
 \label{SCBA}
\end{equation}
This equation was studied numerically by Shon and Ando in Ref.\
\onlinecite{ShonAndo}; we treat it below by analytical means.

Weak disorder introduces an exponentially small energy scale
\begin{equation}
 \Gamma_0
  = \Delta e^{-1/\alpha}.
 \label{gamma0}
\end{equation}
At large energies, $\varepsilon \gg \Gamma_0$, we solve the equation
(\ref{SCBA}) by iterations, while at low energies the solution is found in the
form of a series in powers of $\varepsilon$. The resulting self energy
is\cite{footnote1} (upper sign -- retarded, lower sign -- advanced)
\begin{multline}
 \Sigma(\varepsilon) \\
  = \begin{cases}
      \mp i\Gamma_0 - \dfrac{\varepsilon}{\alpha},
        & |\varepsilon| \ll \Gamma_0; \\[10pt]
      -\alpha \varepsilon \log\dfrac{\Delta}{|\varepsilon|}
      \mp \dfrac{i\pi \alpha |\varepsilon|}{2} \left[
          1 + 2\alpha \log\dfrac{\Delta}{|\varepsilon|}
      \right],
        & |\varepsilon| \gg \Gamma_0.
    \end{cases}
 \label{SigmaSCBA}
\end{multline}
Substituting Eq.\ (\ref{SigmaSCBA}) into Eq.\ (\ref{DoS}), we get the density
of states,
\begin{multline}
 \rho_\text{SCBA}(\varepsilon)
  = \frac{2 \bigl| \mathop{\mathrm{Im}} \Sigma(\varepsilon) \bigr|}
         {\pi^2 v_0^2 \alpha} \\
  = \begin{cases}
      \dfrac{2\Gamma_0}{\pi^2 v_0^2 \alpha},
        & \varepsilon \ll \Gamma_0; \\[0.3cm]
      \dfrac{|\varepsilon|}{\pi v_0^2} \left[
        1 + 2\alpha \log \dfrac{\Delta}{|\varepsilon|}
      \right],
        & \varepsilon \gg \Gamma_0.
    \end{cases}
 \label{DoSSCBA}
\end{multline}
At high energies the found density of states is close to its value in clean
graphene, $\rho_0(\varepsilon) = |\varepsilon|/\pi v_0^2$.

To evaluate the SCBA conductivity, we first find the polarization operator
(\ref{Pi}),
\begin{equation}
 \Pi^{RA}(\varepsilon)
  = \frac{1}{4\pi v_0^2 \alpha}\, \frac{\varepsilon}{\varepsilon -
    \mathop{\mathrm{Re}} \Sigma(\varepsilon)}.
\end{equation}
With the help of Eqs.\ (\ref{Drude}) and (\ref{vertex_correct}) we find the
general expression for the SCBA conductivity
\begin{equation}
 \sigma_\text{SCBA}(\varepsilon)
  = \frac{e^2}{\pi^2} \left[
      \frac{\varepsilon}{\alpha_\text{tr}\, \varepsilon
      -\alpha \mathop{\mathrm{Re}} \Sigma(\varepsilon)}
      +1
    \right].
 \label{DrudeSCBA}
\end{equation}
Here the first term comes from the retarded-advanced (RA) sector whereas the
second term (unity) is the contribution of RR and AA correlators. At high
energies ($\varepsilon \gg \Gamma_0$) the SCBA conductivity is governed by the 
RA-term and takes the form
\begin{equation}
 \sigma_\text{SCBA}(\varepsilon)
  \simeq \frac{e^2}{\pi^2 \alpha_\text{tr}} \left[
      1 - \frac{\alpha^2}{\alpha_\text{tr}} \log \frac{\Delta}{|\varepsilon|}
    \right].
 \label{DrudeSCBA-highE}
\end{equation}
The found conductivity shows a logarithmic energy dependence above an
exponentially small energy scale. At the half filling, $\varepsilon\ll\Gamma_0$,
the SCBA yields $\sigma_\text{SCBA} = 2e^2/\pi^2\hbar$. This value of
conductivity includes contribution of the form $\langle j G^R j G^R\rangle$ and
$\langle j G^A j G^A\rangle$, which were discarded at $\varepsilon \gg
\Gamma_0$. A conductivity value of the order of $e^2/h$ does not make much sense
in the present context, in view of the localization effects. We will return to
this issue in Sec.\ \ref{s4}.

\subsubsection{Logarithmic corrections and renormalization group}
\label{s3.1.2}

\begin{figure}
\centerline{\includegraphics{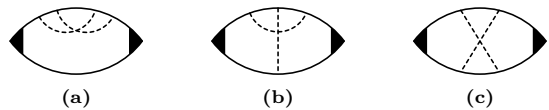}}
\caption{Diagrams yielding logarithmic corrections to the conductivity not
included in the SCBA.}
\label{Fig:LogCorrect}
\end{figure}

The leading term in the Drude conductivity (\ref{DrudeSCBA}) is proportional to
$\alpha_\text{tr}^{-1}$ and is independent of energy. The SCBA gives also
logarithmic corrections, which are small at large energies. There exist,
however, other contributions of the same order that are not included in the SCBA
calculation, see Fig.\ \ref{Fig:LogCorrect}. An efficient tool for resummation
of the logarithmic contributions in all orders is the renormalization group
(RG). For the case of 2D Dirac fermions subjected to various types of disorder
it was developed by Dotsenko and Dotsenko\cite{Dotsenko83} for the random bond
Ising model, by Ludwig \textit{et al}\cite{Ludwig} in the context of the
quantum Hall effect, by Nersesyan \textit{et al}\cite{NersesyanTsvelik} and
Bocquet \textit{et al}\cite{Bocquet} in application to dirty superconductors
with unconventional pairing (see also the review by Altland \textit{et
al}\cite{AltlandSimonsZirnbauer}), as well as by Guruswamy \textit{et
al}\cite{Guruswamy} for a model with chiral disorder ($C_z$ in our notation).
Very recently, Aleiner and Efetov\cite{AleinerEfetov} returned to such a RG in
the context of disordered graphene. The renormalization of the conductivity
gives rise to its dependence on energy (or, equivalently, on the electronic
concentration, see below).

Let us briefly analyze the leading logarithmic corrections and the RG results
and compare them to the SCBA in the simplest case of diagonal disorder with the
only parameter $\alpha = 2\alpha_\text{tr} = \alpha_0$ [see Table
\ref{Tab:sym}]. The diagrams of Fig.\ \ref{Fig:LogCorrect} give logarithmic
corrections proportional to $\alpha_0 \log(\Delta/|\varepsilon|)$ and missed by
SCBA
\begin{equation}
 \delta\sigma
  = \frac{2 e^2}{\pi^2 \alpha_0}\times \begin{cases}
      +2\alpha_0 \log(\Delta/|\varepsilon|), & \text{(a)}; \\
      -2\alpha_0 \log(\Delta/|\varepsilon|), & \text{(b)}; \\
    \end{cases}
 \label{non-scba-long}
\end{equation}
A contribution from the diagram (c), which is potentially of the same order,
vanishes after the angular integration. Since the two contributions in
(\ref{non-scba-long}) cancel each other, the SCBA turns out to give the leading
logarithmic correction even with a correct numerical coefficient,
\begin{equation}
 \sigma(\varepsilon)
  = \frac{2e^2}{\pi^2 \alpha_0} \left[
      1 - 2\alpha_0 \log \dfrac{\Delta}{|\varepsilon|} + \ldots
    \right].
 \label{sigma-leading-log}
\end{equation}
This coincidence in the numerical coefficient seems to be accidental, however.
If one takes into account disorder amplitudes other than $\alpha_0$, the
numerical coefficient in front of the leading logarithmic correction in SCBA
becomes in general different from the correct one (given by RG).

With lowering energy, consideration of the first-order logarithmic correction
(\ref{sigma-leading-log}) becomes insufficient.  As have been already mentioned,
all logarithmic corrections to the density of states and conductivity can be
summed up with the help of the RG, Refs.\ \onlinecite{Dotsenko83, Ludwig,
NersesyanTsvelik, AltlandSimonsZirnbauer, Bocquet, Guruswamy, AleinerEfetov}.
Below we briefly present the results in the simplest case of long-range
Gaussian disorder. For short-range disorder, the consideration is similar but
five running couplings (first 5 in Table \ref{Tab:sym}) characterizing the
disorder should be taken into account. As found in Ref.\
\onlinecite{AleinerEfetov}, this does not affect qualitatively the behavior of
the conductivity.

\begin{figure}
\centerline{\includegraphics{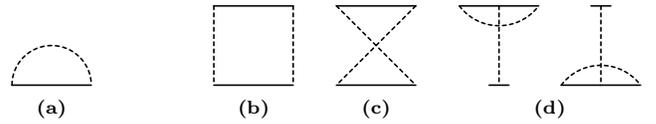}}
\caption{One-loop RG diagrams responsible for the renormalization of (a) the
energy and (b, c, d) the disorder couplings.}
\label{Fig:RG}
\end{figure}

After the disorder averaging, the action for electrons in graphene with
long-range disorder reads
\begin{equation}
 S[\psi]
  = \int d^2 r \Bigl[
      i \bar\psi \bigl(
        \varepsilon + i v_0 \tau_3 \bm{\sigma} \nabla - i0 \Lambda
      \bigr) \psi + \pi v_0^2 \alpha_0 (\bar\psi \psi)^2
    \Bigr].
 \label{action}
\end{equation}
The vector superfield $\psi$ contains $4\times 2\times 2 = 16$ components: the
four-dimensional structure of the one-particle Hamiltonian is complemented by
the advanced--retarded (AR) and the supersymmetric (boson--fermion) structures.
The latter serves to perform the disorder averaging; alternatively, one can use
the replica trick. Further, $\Lambda$ is the third Pauli matrix in the AR
space, and the conjugated field is $\bar\psi = \psi^+ \Lambda$. Under
renormalization, the energy and the disorder strength become running couplings,
$\tilde{\varepsilon} = \varepsilon(L)$ and $\tilde\alpha_0=\alpha_0(L)$, where
$L$ is the running ultraviolet cut-off length (measured in units of
$v_0/\Delta$). As usual, after elimination of large momenta, the real-space
coordinates are rescaled to maintain the ultraviolet cut-off $v_0/\Delta$. The
coefficient $v_0$ of the kinetic term is kept fixed by the field
renormalization (absent in the one-loop order considered below). The relevant
one-loop diagrams are shown in Fig.\ \ref{Fig:RG} (the first two diagrams
renormalizing $\alpha_0$ cancel exactly if the disorder is long-range); the
resulting RG equations read
\begin{align}
 \frac{d\tilde\alpha_0}{d \log L}
  &= 2\tilde\alpha_0^2, \label{eRGeq-alpha} \\
 \frac{d\tilde\varepsilon}{d \log L}
  &= (1 + \tilde\alpha_0) \tilde\varepsilon.
 \label{eRGeq-epsilon}
\end{align}
Note that these equations are different from those for the random mass problem,
Refs.\ \onlinecite{Dotsenko83,Ludwig,Bocquet}, only by a sign in Eq.\
(\ref{eRGeq-alpha}). The RG equation (\ref{eRGeq-alpha}) for the random scalar
potential problem can be found, e.g., in Ref.\ \onlinecite{NersesyanTsvelik}.

Solving these differential equations, we get
\begin{align}
 \tilde\alpha_0
  &= \frac{\alpha_0}{1 - 2\alpha_0 \log L}, \\
 \tilde\varepsilon
  &= \frac{\varepsilon L}{\sqrt{1 - 2\alpha_0 \log L}}.
 \label{eRG}
\end{align}
The renormalization proceeds until the renormalized energy $\tilde\varepsilon$
reaches the value of the cut-off $\Delta$. Using this condition, we find the
value of $L$ at which the RG stops,
\begin{equation}
 L
  = \frac{\Delta}{|\varepsilon|}
    \sqrt{1 - 2\alpha_0 \log \frac{\Delta}{|\varepsilon|}}.
 \label{RGstop}
\end{equation}
The density of states $\rho$ scales as $\varepsilon^{-1}L^2$, i.e.
$\tilde\rho \tilde\varepsilon L^{-2} = \rho \varepsilon$. Thus, according to
Eq.\ (\ref{eRG}), its renormalized value $\tilde\rho$ is
\begin{equation}
 \tilde\rho
  = \rho L \sqrt{1 - 2\alpha_0 \log L}.
 \label{rhoRG}
\end{equation}
At the end point of the RG we have $\tilde\rho = \Delta/\pi v_0^2$. Extracting
$\rho$ from Eq.\ (\ref{rhoRG}) and substituting $L$ from Eq.\ (\ref{RGstop}),
we find the density of states
\begin{equation}
 \rho(\varepsilon)
  = \dfrac{|\varepsilon|}{\pi v_0^2} \left[
      1 - 2\alpha_0 \log \dfrac{\Delta}{|\varepsilon|}
    \right]^{-1}.
 \label{rg-dos}
\end{equation}

Further, the conductivity is determined\cite{footnote-current} by the
renormalized dimensionless strength of the disorder $\tilde\alpha_0$,
\begin{equation}
\label{rg-conductivity}
 \sigma(\varepsilon) = \frac{2 e^2}{\pi^2 \tilde\alpha_0}
  = \frac{2 e^2}{\pi^2 \alpha_0} \left[
      1 - 2\alpha_0 \log \dfrac{\Delta}{|\varepsilon|}
    \right],
\end{equation}
in agreement with Ref.\ \onlinecite{AleinerEfetov}.

We see that the result of the SCBA, Sec.\ \ref{s3.1.1}, agrees qualitatively
with the fully controllable (RG) solution: the conductivity decreases
logarithmically up to an exponentially small scale. The SCBA fails, however, to
give a correct numerical coefficient in the exponent of Eq.\ (\ref{gamma0});
the correct low-energy scale $\Gamma$ is
\begin{equation}
 \Gamma
  = \Delta e^{-1/2\alpha_0}.
\end{equation}
Below this new energy scale, the density of states saturates at a finite value,
and the Drude conductivity (with localization effects discarded) is of the
order of $e^2/h$. Both these important features are correctly reproduced by the
SCBA.

In the experiment, one changes the chemical potential $\mu$ by varying the gate
voltage $V_g$. The electron concentration $n_e$ is proportional to $V_g$, $en_e
= C V_g$, where $C$ is the corresponding capacitance per unit area. Therefore,
the experimentally measured dependence $\sigma(V_g)$ is essentially
$\sigma(n_e)$, up to a simple rescaling. To compare the theory with the
experiment, we find the density
\begin{equation}
 n_e(\mu)
  = 2 \int_0^\mu d\varepsilon\; \rho(\varepsilon)
  \simeq \frac{\mu |\mu|}{\pi v_0^2} \frac{1}{1 - 2\alpha_0 \log \Delta/|\mu|}.
 \label{density-born}
\end{equation}
Combining this with (\ref{rg-conductivity}), we get
\begin{equation}
 \sigma(n_e)
  = \frac{2e^2}{\pi^2\alpha_0} \left(
      1 - \alpha_0\log \frac{\Delta^2}{v_0^2 |n_e|}
    \right).
 \label{sigma-density-born}
\end{equation}

We see that the dependence of conductivity on electron density is only
logarithmic, which should be contrasted with a much stronger, approximately
linear, dependence observed in the experiments.\cite{Novoselov05, Zhang} As we
will see in Sec.\ \ref{s3.3}, such a strong dependence does arise theoretically
in the limit of strong scatterers.

One can use a more general RG approach in the case of generic Gaussian disorder
when all 9 parameters from Table \ref{Tab:sym} are non-zero. In doing so, one
has to calculate the diagrams from Fig.\ \ref{Fig:RG} with all possible
matrices at the vertices of impurity lines. The full set of one-loop
perturbative RG equations can be found in Appendix \ref{App:RG}.

We note that the conductivity calculated above is not the total conductivity
far from the degeneracy point. There are also weak-localization corrections to
the conductivity, which are small for strong enough dephasing or for small
enough systems. As shown in Ref.\ \onlinecite{McCannCondMat06}, it is
convenient to decompose the retarded-advanced Cooperons in the singlet/triplet
representation in both $\Sigma$ and $\Lambda$ channels.  Then only singlets
with respect to $\Sigma$ matter. The general expression for the
weak-localization correction valid for arbitrary disorder then reads
\begin{equation}
 \delta\sigma_{\rm WL}
  = -\frac{e^2}{\pi^2} \log\left(\frac{L_\text{IR}}{l}\right)\,
    [c_0 - 2c_\perp - c_z],
\label{WL}
\end{equation}
where $l$ is the electron mean free path determined by the renormalized
disorder and density of states, and $L_\text{IR}$ is the infrared cutoff set by
either the system size or the dephasing length. In Eq.\ (\ref{WL}), one has to
put $c_i=1$ if disorder preserves the TR-invariance $T_i$ and $c_i=0$ (meaning
that the corresponding Cooperon modes are gapful) otherwise (see Table
\ref{Tab:sym}). For a combination of several disorder types, only those
Cooperon modes remain gapless that correspond to the TR-symmetries preserved
simultaneously by \emph{all} disorder matrices involved. In particular, for the
diagonal disorder $\alpha_0$ all $c_i=1$ which yields antilocalization,
whereas, e.g., for the combination of $\gamma_\perp$ and $\gamma_z$ disorders 
we have $c_0=c_z=1$ and $c_\perp=0$ leading to the absence of the one-loop
correction. On the other hand, for the combination of, e.g., $\beta_z$ and
$\gamma_z$ disorders, $c_0 = 1$, $c_z = c_\perp = 0$, and we get localization.
Note that the weak-localization correction is universal and depends only on the
symmetry of the Hamiltonian.\cite{WB}

\subsection{Unitary limit}
\label{s3.3}

In Sec.\ \ref{s3.1} we have analyzed the behavior of the density of states and
the conductivity in the case when impurities are weak, so that the disorder can
be considered as Gaussian. (In terms of the action, Eq.\ (\ref{action}), this
amounted to keeping, after the ensemble averaging, only the
$(\bar{\psi}\psi)^2$ term and neglecting all higher-order couplings). In this
subsection, we will consider the opposite case, when the electron is strongly
scattered by an impurity and one has to deal with the complete $T$-matrix
(\ref{T}). The analysis of the location of the ``phase boundary'' between the
domains of weak and strong scatterers in the space of microscopic parameters of
the problem is postponed to Sec.\ \ref{s3.4}.

We proceed by first analyzing the results in the framework of the SCTMA, Sec.\
\ref{s3}, and then discuss its accuracy and limitations. Like in the
weak-scatterer limit, the SCTMA can be simplified in the limit of strong
scatterers. Specifically, at large $U$ we neglect unity in comparison with
$U g(\varepsilon)$ in the denominator of Eq.\ (\ref{SCTMA}) and obtain the
following self-consistency equation
\begin{equation}
 \Sigma(\varepsilon)
  = \frac{\eta \Delta^2}
         {\bigl[\varepsilon - \Sigma(\varepsilon)\bigr]
           \log \frac{- \Delta^2}{[\varepsilon - \Sigma(\varepsilon)]^2}
         }.
 \label{SCUA}
\end{equation}
The parameter $\eta$ is the dimensionless concentration of impurities defined
as
\begin{equation}
 \eta
  = \frac{\pi n_\text{imp} v_0^2}{\Delta^2}
    \times \begin{cases}
      1, & \text{short-range}; \\[10pt]
      4, & \text{long-range}.
    \end{cases}
 \label{beta}
\end{equation}
The scattering amplitude $U$ does not enter Eq.\ (\ref{SCUA}). This means the
impurities are effectively considered infinitely strong; the limit that we will
term the \emph{self-consistent unitary approximation} (SCUA). For this type of
impurities, the weak disorder assumption means that their concentration is
small, $\eta \ll 1$.

The characteristic energy scale in the unitary limit is set by the value of
$\Sigma$ at zero energy: $\Sigma(\varepsilon = 0) = \mp i\Gamma_\eta$, which
we find to be
\begin{equation}
 \Gamma_\eta
  \simeq \Delta \sqrt{\frac{\eta}{\log (1/\eta)}}.
\end{equation}
In contrast to its Born-limit counterpart $\Gamma$, which is exponentially
small for $\alpha\ll 1$ , the energy scale $\Gamma_\eta$ depends on the
disorder strength $\eta$ in the power-law fashion. As we see below, this is
intimately connected with a qualitatively different dependence of conductivity
on the Fermi energy.

The further analysis of Eq.\ (\ref{SCUA}) can be performed in the way analogous
to our treatment of the Born limit, Eq.\ (\ref{SCBA}). We get (see Ref.\
\onlinecite{footnote1} concerning the crossover between high- and low-energy
regimes)
\begin{equation}
 \Sigma(\varepsilon)
  = \begin{cases}
      \mp i \Gamma_\eta + \dfrac{\eta \Delta^2 - 2 \Gamma_\eta^2}
        {2(\eta \Delta^2 - \Gamma_\eta^2)}\,\varepsilon,
        & \varepsilon \ll \Gamma_\eta; \\
      \dfrac{\eta \Delta^2}{2\varepsilon} \left[
        \dfrac{1}{\log(\Delta/|\varepsilon|)}
        \mp \dfrac{i\pi \mathop{\mathrm{sgn}}\varepsilon}
            {2\log^2(\Delta/|\varepsilon|)}
      \right],
        & \varepsilon \gg \Gamma_\eta.
    \end{cases}
 \label{SigmaSCUA}
\end{equation}
Here upper (lower) signs correspond to the retarded (advanced) self energy.

Using Eq.\ (\ref{SigmaSCUA}) and the relation between the Green function and
the self energy in the unitary limit, $g = - \eta\Delta^2/4\pi v_0^2\Sigma$,
we get for the density of states, Eq.\ (\ref{DoS}),
\begin{equation}
 \rho_\text{SCUA}(\varepsilon)
  = \frac{\eta \Delta^2}{\pi^2 v_0^2}
    |\mathop{\mathrm{Im}} \Sigma^{-1}(\varepsilon)| \\
  = \begin{cases}
      \dfrac{\eta \Delta^2}{\pi^2 v_0^2 \Gamma_\eta},
        & \varepsilon \ll \Gamma_\eta; \\[0.3cm]
      \dfrac{|\varepsilon|}{\pi v_0^2},
        & \varepsilon \gg \Gamma_\eta.
    \end{cases}
 \label{dos-unitary}
\end{equation}
The density of states is constant at small energy and shows the linear
dependence characteristic for the clean graphene at high energies. To find the
disorder correction to this result, one has to use a more precise value of the
self energy than that given by Eq.\ (\ref{SigmaSCUA}). After two iterations
of the equation (\ref{SCUA}), the linear-in-$\eta$ contribution to the density
of states is obtained (see details in Appendix \ref{App:unitary_rho}),
\begin{equation}
 \rho^{(1)}_\text{SCUA}(\varepsilon)
  = \frac{|\varepsilon|}{\pi v_0^2} \left[
      1 - \alpha_U(\varepsilon)
    \right].
 \label{dos-unitary-1}
\end{equation}
Here the parameter $\alpha_U$ has the meaning of the inverse dimensionless
conductance [see Eq.\ (\ref{sigma-unitary}) below],
\begin{equation}
 \alpha_U(\varepsilon)
  = \frac{\eta\Delta^2}{2\varepsilon^2 \log^2 (\Delta/|\varepsilon|)}
  \sim \frac{n_\text{imp}\lambda_\varepsilon^2}{\log^2 (\Delta/|\varepsilon|)}.
 \label{alpha_U}
\end{equation}
It is of the order of the squared ratio of the electron wavelength
$\lambda_\varepsilon$ at energy $\varepsilon$ to the distance between
impurities, up to a logarithmic factor. The condition $\varepsilon \gg
\Gamma_\eta$ ensures that the relative correction is small.

To find the correction to the density of states of the second order in $\eta$,
one has to go beyond the self-consistent approximation. The diagrams with
intersecting impurity lines become important in this case as we have already
seen it in the Born limit (Sec.\ \ref{s3.1.2}). A rigorous calculation taking
into account all second-order diagrams is given in Appendix
\ref{App:unitary_rho}. The result is
\begin{equation}
 \rho^{(2)}(\varepsilon)
  = \frac{|\varepsilon|}{\pi v_0^2} \left[
      1 - \alpha_U(\varepsilon)
      -2\alpha_U^2(\varepsilon)
        \log\frac{\Delta}{|\varepsilon|} \log\log\frac{\Delta}{|\varepsilon|}
    \right].
 \label{dos-unitary-2}
\end{equation}

In order to calculate the conductivity, we find the polarization operator, Eq.\
(\ref{Pi}),
\begin{equation}
 \Pi^{RA}(\varepsilon)
  = \frac{1}{4\pi v_0^2}\,
    \frac{\eta\Delta^2}{2\bigl| \Sigma(\varepsilon) \bigr|^2}\,
    \frac{\varepsilon - 2\mathop{\mathrm{Re}} \Sigma(\varepsilon)}
         {\varepsilon - \mathop{\mathrm{Re}} \Sigma(\varepsilon)}.
\end{equation}
Substituting it into Eq.\ (\ref{Drude}), we obtain for the conductivity at not
too low energy, $\varepsilon\gg \Gamma_\eta$,
\begin{equation}
 \sigma_\text{SCUA}(\varepsilon)
  = \frac{4 e^2 \varepsilon^2}{\pi^2 \eta \Delta^2} \log^2
    \frac{\Delta}{|\varepsilon|}.
 \label{sigma-unitary}
\end{equation}
Equation (\ref{sigma-unitary}) is written for the case of long-range disorder;
if the disorder is short-range, the vertex correction is absent and the
resulting conductivity is twice smaller.

What about the multiple scattering of electrons on complexes of two or more
impurities (described by mutually ``entangled'' $T$-matrices, Appendix
\ref{App:unitary_rho}) that are not included in the SCUA? We remind the reader
that in the Born limit of weak impurities (Sec.\ \ref{s3.1.2}), similar
multiple scattering processes contribute to the dominant (logarithmic) energy
dependence of the conductivity, Eq.\ (\ref{rg-conductivity}). In the unitary
limit, however, the dominant
energy dependence of the conductivity $\sigma(\varepsilon) \propto
1/\alpha_U(\epsilon)$ comes already from the $\varepsilon$-dependence of a
$T$-matrix describing the scattering off a \emph{single} impurity. Therefore,
the logarithmic corrections to the conductivity, analogous to those in the Born
limit [see Eq.\ (\ref{rg-conductivity})], are of minor importance in the
unitary limit.

As in the case of Born-type disorder, Sec.\ \ref{s3.1}, we now convert the
energy dependence of conductivity into its dependence on the electron
concentration $n_e$. We have according to Eq.\ (\ref{dos-unitary}) (for $\mu
\gg \Gamma_\eta$)
\begin{equation}
 n_e (\mu)
  = \frac{\mu |\mu|}{\pi v_0^2},
 \label{density-unitary}
\end{equation}
so that Eq.\ (\ref{sigma-unitary}) yields
\begin{equation}
 \sigma(n_e)
  = \frac{e^2}{4\pi^2} \frac{|n_e|}{n_\text{imp}}
    \log^2 \frac{\Delta^2}{v_0^2 |n_e|}.
 \label{sigma-density-unitary}
\end{equation}
For the short-range disorder, the result is twice larger. In contrast to the
limit of weak (Born) scatterers, the conductivity shows a strong concentration
dependence: it varies linearly with $n_e$, with a logarithmic correction. This
result compares nicely with the experimentally obtained linear behavior of
$\sigma(n_e)$ (or, equivalently, constant mobility).\cite{Novoselov05, Zhang}
This indicates that the dominant scatterers are strong. Equation
(\ref{sigma-density-unitary}) predicts a logarithmic correction to the linear
behavior, which should become more pronounced if the measurement is extended to
larger gate voltages.

One can also calculate the SCUA conductivity at $\varepsilon \ll \Gamma_\eta$.
In this limit, the contributions $\langle j G^R j G^R\rangle$ and $\langle j
G^A j G^A\rangle$ should also be taken into account. The Drude conductivity
then appears to have exactly the same value $\sigma_\text{SCUA} =
2e^2/\pi^2\hbar$ as in the SCBA, Sec.\ \ref{s3.1.1}. Analogously to the SCBA
case, this result is questionable in view of the localization effects neglected
in the Drude formalism. It is important to recall in this context that
infinitely strong impurities are chiral ($C_z$), yielding a divergent density
of states\cite{Gade93}  (cf. Refs.\ \onlinecite{Hirschfeld02, Altland03}) in
this situation. We will return to the conductivity at half filling for a
chiral disorder in Sec.\ \ref{s4}.

\subsection{Phase diagram}
\label{s3.4}

In the preceding subsections, Sec.\ \ref{s3.1} and \ref{s3.3}, we have studied
the limits of weak (Born) and strong (unitary) scatterers, respectively. We
have found that the behavior of the conductivity is essentially different in
the both limits: it depends only logarithmically on energy in the Born limit,
and shows a linear behavior (with a logarithmic correction) in the unitary
limit.  The aim of the present subsection is to construct a ``phase diagram''
that would predict which of these types of behavior is expected for given
characteristics of disorder. (Of course, we do not mean any phases in the
strict sense; there is a smooth crossover between the Born and the unitary
regime).

The unitary limit corresponds to the neglect of unity in comparison with $U
g(\varepsilon)$ in the denominator of Eq.\ (\ref{SCTMA}). In order to see when
this is justified, we use the large-energy expression (\ref{SigmaSCUA}) for
$\Sigma(\varepsilon)$ and compare $U g(\varepsilon)$ with $1$. This yields the
following energy-dependent value of the parameter $\eta$ at the ``phase
boundary'' between the weak-scatterer and strong-scatterer regimes,
\begin{equation}
 \eta_c(\varepsilon)
  \sim \frac{\alpha \varepsilon^2}{\Delta^2}
       \log^2 \frac{\Delta}{|\varepsilon|},
 \label{beta_c}
\end{equation}
or, equivalently, $\alpha_U(\varepsilon) \sim \alpha$. For $\eta \ll
\eta_c(\varepsilon)$, that is $\alpha_U(\varepsilon) \ll \alpha$, the system
is in the unitary limit. 

\begin{figure}
\includegraphics[width=\columnwidth]{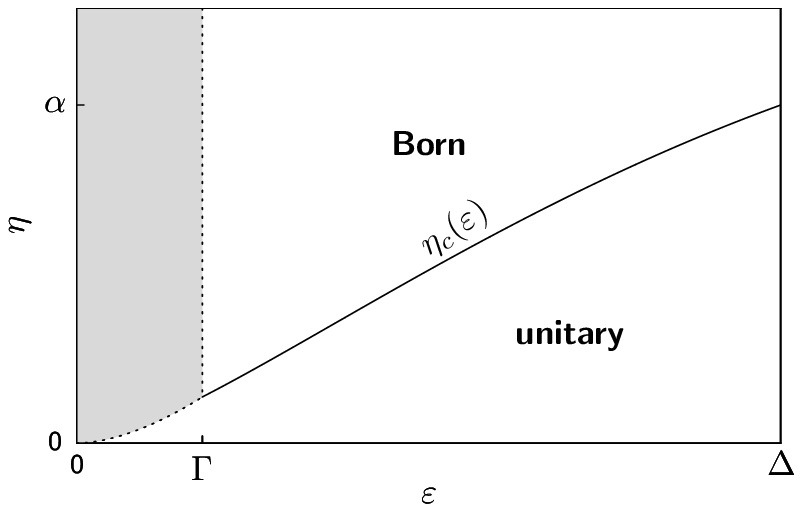} \\[12pt]
\includegraphics[width=\columnwidth]{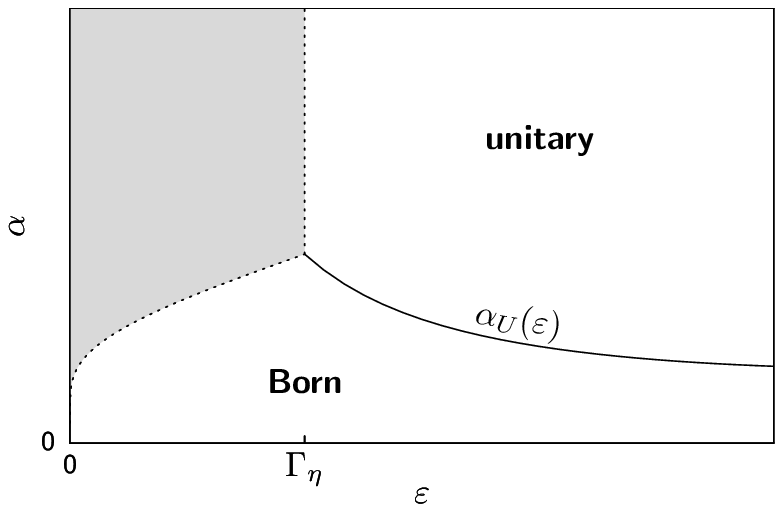}
\caption{``Phase diagram'' in the $\varepsilon$--$\eta$ plane for a fixed
$\alpha$ (upper panel) and in the $\varepsilon$--$\alpha$ plane for a fixed
$\eta$ (lower panel). The solid line is the ``phase boundary'', Eq.\
(\protect\ref{beta_c}) or Eq.\ (\protect\ref{alpha_U}), where the crossover
between the Born and the unitary regimes takes place. At low energies (dashed
part) the density of states saturates, while the Drude conductivity reaches a
value $\sim e^2/h$, implying that generically the localization effects should
become strong.}
\label{Fig:phase}
\end{figure}

In Fig.\ \ref{Fig:phase} we plot the phase diagram in both
$\varepsilon$--$\eta$ and $\varepsilon$--$\alpha$ coordinates. Remarkably, the
system may pass from the Born into the unitary limit when the energy increases
while the disorder remains fixed. In Fig.\ \ref{Fig:phase} this crossover
occurs for a broad range of impurity concentrations,
$\alpha^{-1}\exp(-1/\alpha) \ll \eta \ll \alpha$. At still smaller values of
$\eta$ the system is in the unitary phase at all energies. It is worth
stressing that the unitary phase in Fig.\ \ref{Fig:phase} is established even
for $\alpha \ll 1$, when disorder could be naively considered as weak. The
reason for this effect is as follows. The growth of the density of states with
increasing energy results in a more efficient scattering of higher energy
electrons by an impurity, thus making the scatterer effectively stronger at
higher energies. With increasing $\alpha$ the phase boundary (\ref{beta_c}) in
Fig.\ \ref{Fig:phase} moves upwards, and for $\alpha\agt 1$ the Born phase
disappears altogether.

A unitary-to-Born crossover discussed above would manifest itself in a change
of the behavior of the conductivity, from the linear energy dependence at high
$\varepsilon$ (Sec.\ \ref{s3.1}) to the logarithmic dependence at lower
energies (Sec.\ \ref{s3.3}).\cite{footnote2} Experimentally, the measured
conductivity of graphene shows a linear dependence down to the lowest energy
scale (where $\sigma$ saturates at a value $\simeq 4e^2/h$). This indicates
that the scattering is dominated by strong impurities, which remain in the
unitary part of the phase diagram down to the lowest energies.

\subsection{Charged impurities}
\label{s3.5}

The case of charged impurities deserves a special consideration. If such
impurities are located far from the graphene layer, they are expected to be
screened by the gate and will not be different from finite-range scatterers
considered above. Let us consider, however, charged impurities located near the
graphene layer. The scattering potential of the Coulomb center in 2D is
$V_0(\mathbf{q}) = 2\pi e^2/\chi q$. Taking into account the static screening
by the graphene electron gas in the random phase approximation (RPA), we
obtain\cite{footnote-polarization}
\begin{equation}
 V(\mathbf{q})
  = \frac{2\pi e^2}{\chi q + 2\pi e^2 \rho(\varepsilon)},
 \label{coulomb-screened}
\end{equation}
where $\chi$ is the dielectric constant. Strictly speaking, the RPA is not
justified in graphene since the parameter $r_s = e^2/\hbar v_0 \chi$ is of
order unity. It will be sufficient, however, to find a parametric behavior of
quantities under interest, up to numerical coefficients of order unity.

As follows from Eq.\ (\ref{coulomb-screened}), the intervalley-scattering
component of the Coulomb potential,  $V(k_0) \simeq 2\pi e^2/\chi k_0$ is very
small and can be neglected, so that the Coulomb impurities are of long-range
type. As to the scattering within one valley, it is only slightly anisotropic.
Indeed, the inverse screening length $\kappa = 2\pi e^2\rho(\varepsilon)$ is of
the same order that the characteristic momentum transfer, $\kappa \sim q \sim
\varepsilon/v_0$ for $r_s\sim 1$. Therefore, up to a numerical factor of order
unity, we can neglect $q$ in the denominator of Eq.\ (\ref{coulomb-screened})
(which means a neglect of the anisotropy of the intra-valley scattering). This
brings the screened charged impurities into the class of long-range scatterers
considered above but with an energy-dependent amplitude,
\begin{equation}
 U(\varepsilon)
  = \rho^{-1}(\varepsilon)
  \simeq \frac{\pi v_0^2}{|\varepsilon|}.
 \label{CoulombAmp}
\end{equation}

There is, however, an important difference between a charged impurity and a
long-range potential impurity. The scattering amplitude for slow electrons,
with momenta $q \lesssim |\varepsilon|/v_0$ is given by Eq.\
(\ref{CoulombAmp}), while the electrons with larger momenta are scattered much
less efficiently due to the lack of screening at small distances. This can be
taken into account by setting an effective high-energy cut-off $\Delta \sim
|\varepsilon|$.

Finally, using Eqs.\ (\ref{alpha-long-range}), (\ref{beta}), and (\ref{beta_c})
with $U$ from Eq.\ (\ref{CoulombAmp}) and $\Delta \sim \varepsilon$, we obtain
$\eta_c \sim \eta$. Thus we come to the conclusion that, with charged
impurities, the system is just at the crossover between Born and unitary
regimes. This also justifies the use of the clean density of states in Eq.\
(\ref{CoulombAmp}). Indeed, approaching the crossover from unitary side, the
disorder-induced corrections to $\rho(\varepsilon)$ are negligible, see Eq.\
(\ref{dos-unitary-1}). On the other hand, if one uses the Born expression Eq.\
(\ref{rg-dos}), logarithmic corrections are absent, as long as $\Delta \sim
\varepsilon$, and the clean value of the density of states in Eq.\
(\ref{CoulombAmp}) is again justified. 

The energy and density dependencies of the conductivity of graphene with
Coulomb impurities are thus equivalently given by both Born [Eqs.\
(\ref{rg-conductivity}), (\ref{sigma-density-born}) with energy-dependent
coupling $\alpha_0 \sim n_\text{imp} v_0^2/\varepsilon^2$] and unitary [Eqs.\
(\ref{sigma-unitary}), (\ref{sigma-density-unitary})] expressions with
logarithms omitted,
\begin{equation}
 \sigma
  \sim \frac{e^2 \varepsilon^2}{n_\text{imp} v_0^2}
  \sim \frac{e^2 |n_e|}{n_\text{imp}}.
 \label{conductivity-Coulomb}
\end{equation}

The Born approximation was used for calculating the conductivity in recent
works Refs.\ \onlinecite{Nomura06, Ando06} (see also Ref.\
\onlinecite{Cheianov06Friedel}). The result is consistent with Eq.\
(\ref{conductivity-Coulomb}). A different result (containing an additional
logarithmic factor) was obtained in Ref.\ \onlinecite{Katsnelson-screen}. We
believe that the derivation in Ref.\ \onlinecite{Katsnelson-screen} is
incorrect\cite{footnote-screen} since it employs the quasiclassical
Thomas-Fermi approximation beyond its range of validity (at energies much
larger than $\epsilon_F$).

\section{Conductivity at the degeneracy point: chiral disorder}
\label{s4}

\subsection{Universal conductivity}
\label{s4.1}

In this section we consider the conductivity of graphene at half filling,
$\varepsilon = 0$. The Drude conductivity obtained self-consistently in Sec.\
\ref{s3} in both Born and unitary limit has the value $\sigma = 2 e^2/\pi^2
\hbar$ at this point.  Since this value is of the order of conductance quantum,
this is by no means the end of the story: the localization effects become
strong at half filling. If the intervalley scattering is weak (long-range
disorder potential), an intermediate temperature range exists where the
conductivity correction is positive\cite{Suzuura02, Khveshchenko06loc,
McCannCondMat06, AleinerEfetov} due to the additional Berry phase $\pi$
associated with the electron pseudospin in the sublattice space. This situation
belongs to the symplectic symmetry class. With lowering temperature $T$, the
intervalley scattering comes into play and a crossover to the orthogonal
symmetry class occurs.\cite{McCannCondMat06, AleinerEfetov, Altland06} The
localization correction becomes negative and drives the system into the strong
localization regime.\cite{AleinerEfetov} Thus, for a generic disorder, the
conductivity at half filling should have a pronounced temperature dependence
and get strongly suppressed with lowering $T$. Surprisingly, this is not what
is observed in the experiment. The conductivity has been
found\cite{Novoselov05, Zhang} to be close to the value $4e^2/h$, remaining
$T$-independent in a broad range of temperatures. The aim of this section is to
analyze whether and in what situation this behavior may be expected
theoretically. According to what was said above, this might only happen, if at
all, for a particular type of disorder.

The special class of disorder that we will consider in this section is the
randomness that preserves one of the chiral symmetries (\ref{chiral}) of the
clean graphene Hamiltonian. Some possible realizations of such type of disorder
were listed in Sec.\ \ref{s2.3}. Whether the dominant disorder in graphene may
be of this kind is an open issue, which may be related to technological aspects
of the sample preparation. Our aim here will be to analyze what are
consequences of the assumption of chiral character of disorder.

A peculiar behavior of 2D systems with chiral disorder with respect to
localization effects has been demonstrated by Gade and
Wegner.\cite{GadeWegner91, Gade93} They considered a random hopping problem on
a square lattice and showed that at zero energy, where the system possesses the
chiral symmetry, the RG $\beta$-function of the corresponding $\sigma$-model
vanishes to all orders in the inverse conductivity, implying that the
conductivity is not renormalized. This absence of usual infrared-singular
corrections to the conductivity due to cooperon- and diffuson- loops can be
attributed to the fact that the ``antilocalizing'' interference corrections to
the density of states cancel the localization corrections to the diffusion
coefficient. The density of states has been found\cite{Gade93, Guruswamy,
Motrunich, MudryDOS} to diverge as $\rho(\varepsilon) \sim \varepsilon^{-1}
f(\varepsilon)$ for $\varepsilon \to 0$, where $f(\varepsilon)$ gives the
subleading $\varepsilon$-dependence and provides the convergence of the total
number of electronic states.\cite{footnote3}  At any finite $\varepsilon$ the
chiral symmetry is broken and the localization on the scale $\xi(\varepsilon)
\propto |f(\varepsilon)|^{-1/2}$ occurs.\cite{Gade93} The states at the band
center $\varepsilon=0$ are delocalized and the conductivity
$\sigma(\varepsilon=0)$ takes a finite value depending on the disorder
strength. According to the classification of Refs.\
\onlinecite{Zirnbauer,AltlandZirnbauer,AltlandSimonsZirnbauer}, the system
studied in Refs.\ \onlinecite{GadeWegner91, Gade93} belongs to the chiral
symmetry class AIII.

While the results of Refs.\ \onlinecite{GadeWegner91, Gade93} suggest that one
may expect a finite zero-energy conductivity in our problem, they cannot be
directly applied. Indeed, the dimensionless Drude conductivity at $\varepsilon
= 0$ is of order unity in our case, whereas it should be large to justify the
derivation of the $\sigma$ model and of the perturbative RG. Another related
peculiarity of the problem we are considering is the Dirac dispersion of
carriers. This will allow us to prove below a statement that is still stronger
than that of Gade and Wegner: we will show that for certain types of chiral
disorder all disorder-induced contributions to conductivity cancel.

\subsubsection{$C_0$-chirality: symmetry consideration}
\label{s4.1.1}

Let us consider the disorder which preserves the $C_0$-chirality, $H =
-\sigma_3 H \sigma_3$. The random part of the Hamiltonian  contains matrices
$\sigma_{0}\tau_{3}$, $\sigma_{1,2}\tau_{1,2}$, and $\sigma_{1,2}\tau_0$. 
According to the Table I, in the case of weak disorder, the corresponding
coupling constants are $\alpha_\perp$, $\beta_\perp$, and $\gamma_\perp$.
While the disorder characterized by $\beta_\perp$ and $\gamma_\perp$
preserves the time-reversal invariance $T_0$, the $\alpha_\perp$-disorder,
being physically a random vector potential, violates the $T_0$-symmetry.

According to Ref.\ \onlinecite{BernardLeClair}, the random Dirac Hamiltonians 
preserving the $C_0$-chirality and violating the TR-symmetry (case 1 of
Ref.\ \onlinecite{BernardLeClair}) belong to the chiral symmetry class AIII,
while the combination of $C_0$-chirality and $T_0$-symmetry (case 6 of
Ref.\ \onlinecite{BernardLeClair}) drives the system into the Bogolyubov -- de
Gennes symmetry class CI. In both cases, the low-energy theory ($\sigma$
model) is affected by the presence of the Wess-Zumino-Witten term in the
action. 

The one-loop RG equations for $C_0$-disorder read (see Appendix \ref{App:RG}):
\begin{align}
 \frac{\partial{\alpha_\perp}}{\partial \log L}
  &= 0, \label{RGC0alphaperp} \\
 \frac{\partial{\beta_\perp}}{\partial \log L}
  &= 4 \beta_\perp \gamma_\perp, \label{RGC0betaperp} \\
 \frac{\partial{\gamma_\perp}}{\partial \log L}
  &= \beta_\perp^2. \label{RGC0gammaperp}
\end{align}
Note that the equation (\ref{RGC0alphaperp}) for $\alpha_\perp$ is split from 
Eqs.\ (\ref{RGC0betaperp}), (\ref{RGC0gammaperp}). This set of equations is
identical to Eq.\ (19) of Ref.\ \onlinecite{AltlandSimonsZirnbauer} with $g' =
\alpha_\perp$, $g = 2\gamma_\perp$, $g_{\pi 0} = 0$, and $g_{\pi \pi} =
2\beta_\perp$. In Ref.\ \onlinecite{AltlandSimonsZirnbauer}, these couplings
described the scattering between the four nodal points of the spectrum of
disordered $d$-wave superconductor. Our problem with only two nodes corresponds
to setting the coupling between the neighboring nodes in $d$-wave
superconductor to zero, $g_{\pi 0} = 0$, while retaining the forward-scattering
(intranode, $g$) and backscattering (scattering between the opposite nodes,
$g_{\pi \pi}$) amplitudes. This situation (non-Abelian vector potential
problem) was considered in Ref.\ \onlinecite{NersesyanTsvelik} (see also Refs.\
\onlinecite{Fendley,LudwigCM}), where the density of states was shown to vanish
in the limit $\varepsilon\to 0$ as 
\begin{equation}
 \rho(\varepsilon)
  \propto |\varepsilon|^{1/7}.
\end{equation} 
Here $1/7 = 1/(2N^2-1)$, where $N=2$ is the number of flavors (nodes).

In the presence of the random vector potential only ($\alpha_\perp$ coupling,
preserving all the four chiralities simultaneously, class AIII), the density of
states also goes to zero with decreasing energy, but with a non-universal
exponent\cite{NersesyanTsvelik, Ludwig} which depends on $\alpha_\perp$:
\begin{equation}
 \rho(\varepsilon)
  \propto |\varepsilon|^{(1-\alpha_\perp)/(1+\alpha_\perp)}.
\label{rhoA}
\end{equation} 
Note that in this random vector potential problem, the disorder strength
remains non-renormalized, see Eq.\ (\ref{RGC0alphaperp}) [in fact, the one-loop
equations for $\alpha_\perp$ and $\varepsilon$ are exact, see Refs.\
\onlinecite{Ludwig,Guruswamy}]. Therefore, $\alpha_\perp$ does not generate the
scale $\Gamma$ and the one-loop result for the density of states, Eq.\
(\ref{rhoA}), holds in the whole range of energies below $\Delta$.

\subsubsection{$C_0$-chirality: conductivity at Dirac point}
\label{s4.1.2}

We are now going to study the conductivity in the situation when disorder
preserves $C_0$-chirality. The chiral symmetry $C_0$ allows one to relate
retarded and advanced Green functions:
\begin{equation}
 \sigma_3 G^{R(A)}(\varepsilon; \mathbf{r},\mathbf{r'}) \sigma_3
  = - G^{A(R)}(-\varepsilon; \mathbf{r},\mathbf{r'}).
 \label{GRA}
\end{equation}
The conductivity is given by the Kubo formula (\ref{Kubo}), which we rewrite
here in the full form,
\begin{multline}
 \sigma^{xx}
  = \frac 1{\pi} \int d^2(r-r') \mathop{\mathrm{Tr}} \Bigl[
      j^x G^R(0,\mathbf{r},\mathbf{r'})
      j^x G^A(0,\mathbf{r'},\mathbf{r})
    \\
      -\frac{1}{2} j^x G^R(0,\mathbf{r},\mathbf{r'})
      j^x G^R(0,\mathbf{r'},\mathbf{r})
    \\
      -\frac{1}{2} j^x G^A(0,\mathbf{r},\mathbf{r'})
      j^x G^A(0,\mathbf{r'},\mathbf{r})
    \Bigr].
 \label{KuboFull}
\end{multline}
Now we use the identity (\ref{GRA}) to trade all advanced Green functions in
Eq.\ (\ref{KuboFull}) for retarded ones and thus to present the conductivity in
terms of retarded Green functions only. Further, we exploit the following
important relation between the components of the current operator
(\ref{current}), 
\begin{equation}
 \sigma_3 j^x
  = -j^x \sigma_3
  = i j^y,
 \label{current-x-y}
\end{equation}
which is the consequence of the Dirac spectrum. At this point, our problem
differs from that considered by Gade and Wegner\cite{GadeWegner91,Gade93} who
dealt with a bipartite \emph{square} lattice with a non-linear electronic
spectrum.

The transformations Eqs.\ (\ref{GRA}) and (\ref{current-x-y}) allow us to cast
the Kubo formula in the following form:
\begin{multline}
 \sigma^{xx}
  = -\frac 1{\pi} \sum_{\alpha = x,y} \int d^2(r-r')
    \\
    \times \mathop{\mathrm{Tr}} \Bigl[
      j^\alpha G^R(0;\mathbf{r},\mathbf{r'})
      j^\alpha G^R(0;\mathbf{r'},\mathbf{r})
    \Bigr].
 \label{KuboRR}
\end{multline}
At first glance, this expression is zero due to the gauge invariance. Indeed,
the right-hand side of Eq.\ (\ref{KuboRR}) is proportional to the second
derivative of the partition function $Z[\mathbf{A}] = \mathop{\mathrm{Tr}}\log
G^R[\mathbf{A}]$ (or, equivalently, first derivative of the current
$\mathop{\mathrm{Tr}} j^\alpha G^R[\mathbf{A}]$) with respect to the constant
vector potential $\mathbf{A}$. The gauge invariance implies that a constant
vector potential does not affect gauge-invariant quantities like the partition
function or the current, so that the derivative is zero. This argument is,
however, not fully correct, in view of a quantum anomaly present in this
problem. The elimination of $\mathbf{A}$ amounts technically to a shift in the
momentum space $\mathbf{k} \to \mathbf{k} - e\mathbf{A}$, which naively does
not change the momentum integral. If we consider a formal expansion in the
disorder strength, this argument will indeed hold for all terms involving
disorder but not for the zero-order contribution. The momentum integral $\int
d^2 k \mathop{\mathrm{Tr}} j^\alpha G^R_0(\mathbf{k})$ is
ultraviolet-divergent and the shift of variable is illegitimate. This anomaly
was first identified by Schwinger\cite{Schwinger} for $1+1$-dimensional
massless Dirac fermions. In the Schwinger model, the polarization operator is
not affected  by an arbitrary external vector potential $\mathbf{A}(x,t)$ and
is given by the anomalous contribution, yielding a photon mass in the $1+1$
electrodynamics.\cite{Schwinger, Peskin} In our analysis, the role of
$\mathbf{A}(x,t)$ is played by the chiral disorder. The explicit calculation of
the zero-order diagram (the one with no disorder included) yields
\begin{equation}
 \sigma
  = -\frac{8 e^2 v_0^2}{\pi} \int \frac{d^2k}{(2\pi)^2}\,
    \frac{\delta^2}{(v_0^2 k^2 + \delta^2)^2}
  = \frac{2e^2}{\pi^2}.
 \label{universal}
\end{equation}
Here $\delta$ is an infinitesimal imaginary part in the denominator of the
Green function; we will return to its role and physical meaning below. We
note that the same universal value of the conductivity in the situation when
the only type of disorder is the abelian random vector potential
($\alpha_\perp$) was previously obtained in Ref.\ \onlinecite{Ludwig}.

An alternative derivation of the same result is based on the Ward identity
\begin{equation}
 -i e (\mathbf{r}-\mathbf{r'}) G^R(0; \mathbf{r}, \mathbf{r'})
  = [G^R \mathbf{j} G^R] (0; \mathbf{r},\mathbf{r'}).
\end{equation}
Averaging it over disorder, plugging it into Eq.\ (\ref{KuboRR}), transforming
to the momentum space, and performing the integration by parts, we are left
with
the surface contribution only,
\begin{equation}
 \sigma
  = -\frac{e v_0}{4 \pi^3} \oint d\mathbf{k}_n \mathop{\mathrm{Tr}} \bigl[
      \mathbf{j} G^R(\mathbf{k})
    \bigr],
\end{equation}
where the integral is taken over a large circle $|{\mathbf{k}}| =
\mathrm{const} \to \infty $. For large momenta the Green function can be
replaced by its bare value, which yields again the universal conductivity
\begin{equation}
 \sigma
  = \frac{e^2}{\pi^3} \oint \frac{d\mathbf{k}_n \mathbf{k}}{k^2}
  = \frac{2 e^2}{\pi^2}.
\label{universal-2}
\end{equation}
This universal value of the conductivity is independent of the ultraviolet
cut-off in the momentum space. This signifies that the integral in Eq.\
(\ref{KuboRR}) is accumulated in the vicinity of the degeneracy point, as seen
explicitly in Eq.\ (\ref{universal}). The fact that, in a realistic system, the
linearization of the spectrum ceases to be valid at high momenta does not spoil
the derivation:  the functions $G^R$ and $G^A$ are essentially equal to each
other there, so that the integrand of Eq.\ (\ref{KuboFull}) is cancelled.

It is worth emphasizing that, as is clear from the derivation of Eq.\
(\ref{universal-2}), it assumes that the ultraviolet cut-off $\Delta$ is much
larger than the disorder-induced energy scale $\Gamma$. (More accurately, here 
$\Gamma$ is the low-energy electron relaxation rate determined as a scale where
the dimensionless Drude conductivity is of order of unity, or, equivalently,
where the RG flow enters the strong coupling regime.) In other words, the
disorder is weak, i.e. $\alpha \ll 1$ for Gaussian disorder.  One more
formulation of this condition is that for energies comparable to the cut-off,
$\varepsilon \sim \Delta$, the Drude conductivity considered in Sec.\ \ref{s3}
is large (compared to $e^2/h$). This condition, that we assume throughout the
paper, is very well fulfilled in the experiments.\cite{Novoselov05, Zhang}
Violation of this condition would imply that the disorder is so strong that it
completely destroys the Dirac character of the spectrum. In this situation the
universal value of the conductivity (\ref{universal}), (\ref{universal-2}) of
the chiral-symmetric system would not survive. The corrections to the universal
value of the conductivity are exponentially small: $\delta\sigma \lesssim
e^2 \Gamma/\Delta$, which implies that there are no corrections to any order in
the perturbative expansion of $\sigma(\varepsilon = 0)$ in $\alpha\ll 1$.

The above derivation of the universal conductivity remains valid for the case
when a magnetic field of an arbitrary strength is applied: the vector potential
$A_\alpha$ couples to the current, i.e. to the matrices $\sigma_\alpha$,
$\alpha=x,y$, thus preserving the chiral symmetry. In this context, it is worth
mentioning the result of Hikami, Shirai, and Wegner\cite{HikamiShiraiWegner93}
who found that the longitudinal conductance in the center of the lowest Landau
level of the chiral-disordered 2D electron gas is equal exactly by $\sigma =
2e^2/\pi^2\hbar$ in the limit of very strong magnetic field, when the Landau
level mixing can be neglected. Their finding can be considered as a
$B\to\infty$ limit of our general result. Indeed, in this limit the kinetic
energy is frozen, so that the difference between the electron dispersion on the
square lattice (considered in Ref.\ \onlinecite{HikamiShiraiWegner93}) and the
graphene lattice becomes immaterial.

We turn now to an important and delicate point related to the above derivation
of the universal conductivity (\ref{universal}). Specifically, we have
introduced an infinitesimally small imaginary part of energy, $\delta$.
Physically, it has  a meaning of the electron lifetime or, alternatively, a
dephasing rate, and can be thought as modelling processes of escape of
electrons in some reservoir or some dephasing mechanism. Models with such a
uniform constant value of $\delta$ were used in the literature to imitate
dephasing in quantum dots, see e.g. Ref.\ \onlinecite{Efetov}.

In our calculation, $\delta$ has served as an infrared regulator for the
theory. Although it has dropped from the final result, its role is not
completely innocent. Depending on the physical situation, the infrared
regularization may be provided by different quantities, which, as we are going
to discuss, will influence the value of the conductivity. Specifically, in
addition to $\delta$, we can imagine the following sources of the infrared
cut-off: (i) finite frequency, (ii) finite system size, and (iii)
interaction-induced dephasing at finite temperature. In Sec.\ \ref{s4.2} we
will analyze the frequency dependence of the conductivity. As to the situations
when the temperature or the system size govern the infrared behavior, we
restrict ourselves to brief comments only, relegating a detailed analysis to
future work.

\subsubsection{$C_z$-chirality}
\label{s4.1.3}

Let us now turn to the disorder which preserves the $C_z$-chirality, $H =
-\sigma_3 \tau_3 H \sigma_3 \tau_3$. The random part of the Hamiltonian may
then contain matrices $\sigma_3 \tau_{1,2}$, $\sigma_{1,2} \tau_3$,
$\sigma_{1,2} \tau_0$, and $\sigma_{0} \tau_{1,2}$. The first two (the
corresponding coupling constants are $\beta_0$ and $\alpha_\perp$) violate the
time-reversal symmetry $T_0$, the last two ($\gamma_\perp$ and $\beta_z$)
preserve it, see Table I. Note that the disorder characterized by
$\alpha_\perp$ and $\gamma_\perp$ (real and imaginary vector potentials,
respectively) also preserves the chiral symmetry $C_0$ considered above.

According to Ref.\ \onlinecite{BernardLeClair}, random Dirac Hamiltonians
preserving the $C_z$-chirality and violating the TR-symmetry (case 2 of Ref.\
\onlinecite{BernardLeClair}) belong to the chiral unitary symmetry class AIII.
The combination of $C_z$-chirality and the time reversal invariance $T_0$ (case
$9_+$ of Ref.\ \onlinecite{BernardLeClair}) corresponds the chiral orthogonal
symmetry class BDI. Finally, the combination of $C_z$-chirality and
$T_z$-symmetry (case $9_-$ of Ref.\ \onlinecite{BernardLeClair}) falls into the
chiral symplectic symmetry class CII.

The one-loop RG equations for $C_z$-disorder read (see Appendix \ref{App:RG}):
\begin{align}
 \frac{\partial{\alpha_\perp}}{\partial \log L}
  &= 2\beta_0\beta_z, \label{RGC3alphaperp} \\
 \frac{\partial{\beta_0}}{\partial \log L}
  &= 2\alpha_\perp (\beta_0+\beta_z), \label{RGC3beta0} \\
 \frac{\partial{\beta_z}}{\partial \log L}
  &= 2\alpha_\perp (\beta_0+\beta_z), \label{RGC3betaz} \\
 \frac{\partial{\gamma_\perp}}{\partial \log L}
  &= \beta_0^2+\beta_z^2. \label{RGC3gammaperp}
\end{align}
This model was considered in Ref. \onlinecite{Guruswamy}; the RG equations
(\ref{RGC3alphaperp})--(\ref{RGC3gammaperp}) agree with the set of equations
(4.84) in Ref.\ \onlinecite{Guruswamy} with $g_\mu = \beta_0/\sqrt{2}$, $g_{A'}
= \alpha_\perp$, $g_A = \gamma_\perp$, and $g_m = \beta_z/\sqrt{2}$. If the
system is time reversal ($T_0$) invariant, only the couplings $\beta_z$ and 
$\gamma_z$ survive; this case was considered in Refs.\ \onlinecite{Hatsugai,
Guruswamy}. The density of states in the generic $C_z$-case
diverges\cite{GadeWegner91, Gade93, footnote3, Guruswamy, Motrunich, MudryDOS}
in the limit $\varepsilon\to 0$, see Sec.\ \ref{s4.1} [for the case of the
random vector potential, see Eq.\ (\ref{rhoA})].

Let us turn to the conductivity at half filling for a generic disorder
preserving the $C_z$ chirality. The proof of the universality of the
conductivity based on gauge-invariance argument does not work now. Indeed, the
$C_z$-chirality transformation of the Green's function
\begin{equation}
 \sigma_3\tau_3 G^{R(A)}(\varepsilon; \mathbf{r},\mathbf{r'}) \sigma_3\tau_3
  = - G^{A(R)}(-\varepsilon; \mathbf{r},\mathbf{r'})
\end{equation} 
generates the new vector vertices $j^{x,y}\tau_3$ instead of currents (these new 
vertices can be considered as $\mathbf{j}_5$ currents of Dirac fermions). Then
we are left with the $G^RG^R$-type correlators of both $j^{x,y}$ and
$j^{x,y}_5$. The latter can not be obtained as derivatives of the partition
function with respect to the constant vector potential $\mathbf{A}$.
Nevertheless, for weak disorder we find that the conductivity at
half filling is still universal, $\sigma(\varepsilon=0)=2e^2/\pi^2$, up to 
corrections in powers of disorder strength.

To show this, we first calculate the perturbative correction
$\delta\sigma^{(1)}$ to the conductivity of a pure system at the first order in
disorder strength and find that it vanishes, $\delta\sigma^{(1)}=0$. This
implies that the conductivity at $\varepsilon=0$ does not depend on the
ultraviolet cutoff $\Delta$. Indeed, all the contributions generated by the RG
(and thus depending on the ratio $\Delta/\delta$) sum up to zero, because we
can use the fully renormalized disorder as an effective single impurity line in
$\delta\sigma^{(1)}=0$. The second-order perturbative calculation yields
\begin{equation}
 \delta\sigma^{(2)}
  = \frac{e^2}{2\pi^2}(\beta_0 - \beta_z)^2.
\label{deltasigma2}
\end{equation}
We note that the combination $\beta_0 - \beta_z$ is not renormalized during the
RG procedure, as follows from Eqs.\ (\ref{RGC3beta0}) and (\ref{RGC3betaz}). 
This is in agreement with the above RG argument for the first-order correction.
Thus the conductivity at the Dirac point can be presented as a series in the
parameter $\beta_0-\beta_z$. Next, we recall that for $C_z$-chirality, the RG
$\beta$-function of the Gade-Wegner $\sigma$ model\cite{GadeWegner91, Gade93}
vanishes to all orders, so that there are no singular quantum-interference
corrections to $\sigma(\varepsilon=0)$ due to the soft modes (impurity
ladders). This proves that the expansion of $\sigma(\varepsilon=0)$ in powers
of $\beta_0-\beta_z$ converges. Thus for the case of weak disorder the
conductivity is universal with small corrections in powers of the disorder
strength (unlike in the case of the $C_0$-chirality, where the corrections are
nonperturbative in the disorder strength.)
 
\subsubsection{$C_\perp$-chirality}
\label{s4.1.4}

Finally, let us discuss the case of $C_{x,y}$-chirality (couplings
$\alpha_\perp$, $\gamma_0$, and $\gamma_z$). Each of these chiralities taken
separately is similar to the $C_z$-chirality. However, in an isotropic system
considered here, both $C_x$ and $C_y$ chiralities are expected to be present
simultaneously. This implies that the disordered Hamiltonian anticommutes with
both $\tau_1$ and $\tau_2$ and hence is proportional to $\tau_3$. Thus it is
split into two equivalent copies. Therefore, the symmetry of the problem is
governed by the properties of the ``sub-Hamiltonians" and its chirality is in
fact fictitious. In particular, the generic case with all  $\alpha_\perp$,
$\gamma_0$, and $\gamma_z$ present,\cite{Ludwig} corresponds to the
conventional Gaussian unitary class A (Quantum Hall effect). A single coupling
$\gamma_z$ corresponds to the symmetry class D (random mass problem). In all
these cases the system is in a critical phase so one can expect a finite
conductivity at $\varepsilon = 0$. For the sake of completeness we present the
RG equations for the $C_{\perp}$-chirality:
\begin{align}
 \frac{\partial{\alpha_\perp}}{\partial \log L}
  &=4 \gamma_0\gamma_z, \label{RGCxyalphaperp} \\
 \frac{\partial{\gamma_0}}{\partial \log L}
  &= 2 (\alpha_\perp + \gamma_0)(\gamma_0 + \gamma_z), \label{RGCxygamma0} \\
 \frac{\partial{\gamma_z}}{\partial \log L}
  &= 2 (\alpha_\perp - \gamma_z)(\gamma_0 + \gamma_z). \label{RGCxygammaz}
\end{align}
We are not aware of realistic examples of the disorder preserving the
$C_{\perp}$-chirality in the context of the transport in disordered graphene.
Therefore, we will not consider this case in the rest of the paper.

\subsection{Conductivity at finite frequency}
\label{s4.2}

In this subsection, we analyze the frequency dependence of the conductivity.
For completeness, we also keep a small level width $\delta$ introduced above.
It
was crucial for the argument leading to Eq.\ (\ref{universal}) that the system
is exactly at half filling, $\varepsilon = 0$. A non-zero frequency implies an
integration over the energy range of the width $\omega$, which breaks the
chiral symmetry. When the frequency $\omega$ is much smaller than $\delta$,
this effect is however negligible, the infrared regularization is provided by
$\delta$, and the universal result (\ref{universal}) survives. In its turn,
$\delta$ plays no role when $\omega\gg\delta$: it is the frequency that serves
as a dominant infrared cut-off now. In the high-frequency limit, $\omega \gg
\Gamma$, the situation simplifies again: one can neglect the effect of disorder
altogether and calculate the conductivity by a simple Kubo formula with bare
Green functions. The result for the real part of the conductivity is again
universal but with a slightly larger value.\cite{Ludwig, Falkovsky06}
\begin{widetext}
\begin{multline}
 \mathop{\mathrm{Re}} \sigma(\omega \gg \Gamma)
  = \frac{2}{\pi} \int_0^\omega \frac{d\varepsilon}{\omega}
    \int \frac{d^2 k}{(2\pi)^2} \mathop{\mathrm{Tr}} \Bigl[
      j^x \mathop{\mathrm{Im}} G_0^R(\varepsilon - \omega, \mathbf{k})
      j^x \mathop{\mathrm{Im}} G_0^R(\varepsilon, \mathbf{k})
    \Bigr] \\
  = 8 \pi e^2 v_0^2 \int_0^\omega \frac{d\varepsilon}{\omega}
    \int \frac{d^2 k}{(2\pi)^2}
    |\varepsilon - \omega| \delta \bigl[ (\varepsilon - \omega)^2 - k^2 \bigr]
    |\varepsilon| \delta \bigl[ \varepsilon^2 - k^2 \bigr]
  = \frac{e^2}{4}.
\end{multline}
\end{widetext}

A very interesting new situation arises in the intermediate regime, $\delta \ll
\omega \ll \Gamma$. Here $\omega$ plays a twofold role, leading to two
competing effects. On one hand, as discussed above, the frequency drives the
system away from the chiral-symmetric point and thus restores localization. On
the other hand, the frequency cuts off the singular localization correction.
Which of these effects wins? To answer this question, one should compare
$\omega$ with the level spacing in the localization area,
$\Delta_\xi(\varepsilon)$, where $\varepsilon\sim\omega$. In order to find the
scaling of $\Delta_\xi$ with energy, we consider a RG transformation that
drives the system away from the chiral fixed point. The RG stops when the
renormalized energy $\tilde{\varepsilon}$ reaches the macroscopic scale
$\Delta$; on such scales the disorder becomes already strong since the initial 
value of $\varepsilon$ was below $\Gamma$.  In this strongly disordered case, 
the value of the running ultraviolet cut-off length $L v_0/\Delta$
(corresponding to the renormalized electron wavelength) determines then the
localization length $\xi$. As discussed in Sec.\ \ref{s3.1.2}, the density of
states $\rho$ scales as $\varepsilon^{-1} L^2$. Therefore,
\begin{equation}
 \frac{\rho\varepsilon}{(v_0/\Delta)^2}
  \sim \frac{\tilde{\rho}\tilde{\varepsilon}}{\xi^2},
\end{equation}
implying for the level spacing at the length $\xi$,
\begin{equation}
 \Delta_\xi(\varepsilon)
  \equiv \frac{1}{\rho(\varepsilon)\xi^2(\varepsilon)}
  \sim \varepsilon
  \sim \omega.
 \label{nichja}
\end{equation}
This result is rather general and is only based on the fact that the operator
governing the flow of the system away from  criticality couples to the energy
in the action. One can of course explicitly verify that the results of
Ref. \onlinecite{Gade93} for the density of states and the localization length
quoted in Sec.\ \ref{s4.1} satisfy Eq.\ (\ref{nichja}).

We conclude that the two competing effects of the frequency (the localization
and the infrared regularization) ``make a draw'' -- both of them are equally
important. Therefore, the system turns out to be, roughly speaking, half way
between the chiral fixed point and the conventional symmetry. This results in a
new universal (frequency-independent) value of the conductivity $\sigma_\omega
\sim e^2/h$ in the considered regime $\delta \ll \omega \ll \Gamma$. More
precisely, this value depends on the type of chirality and the symmetry class
of the system away from the degeneracy point. In particular, the system
with generic $C_0$ and $C_z$ chiral disorder with (without) TR symmetry $T_0$
is driven into the Wigner-Dyson orthogonal (respectively, unitary) symmetry
class by finite energy. On the other hand, the system with $C_z$- and
$T_z$-invariant disorder ($\beta_0$ and $\gamma_\perp$) falls into the Gaussian
symplectic symmetry class away from $\varepsilon = 0$.

\begin{figure}
\includegraphics[width=\columnwidth]{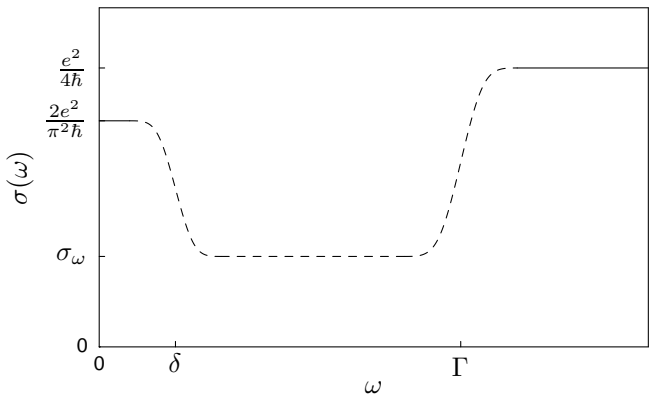}
\caption{Frequency dependence of conductivity in a system with chiral
disorder. At intermediate frequency, $\delta \ll \omega \ll \Gamma$, the
conductivity acquires some universal value $\sigma_\omega$ of the order of
$e^2/h$ which is not known analytically. This value depends on the type of
chirality and the symmetry class of the system away from the degeneracy point.}
\label{Fig:SigmaOmega}
\end{figure}

The frequency dependence of the conductivity is sketched in Fig.\
\ref{Fig:SigmaOmega}. Despite its universality (for a given symmetry), the
value $\sigma_{\omega}$  most likely cannot be calculated analytically, since
this would require an exact knowledge of the full crossover between the chiral
and the normal classes.

\subsection{Additional comments}
\label{s4.3}

In Secs.\ \ref{s4.1} and \ref{s4.2} we have analyzed the conductivity in the
case when the dominant infrared regularization is provided either by the
inverse life time $\delta$ or by the frequency $\omega$. As has been mentioned
above, this role may be alternatively played by the interaction-induced
dephasing at finite temperature or by the finite size of the system. Leaving a
detailed analysis of these problems for the future, we only make some comments
on them in Secs.\ \ref{s4.3.1} and \ref{s4.3.2} below. Finally, in Sec.\
\ref{s4.3.3} we briefly discuss what happens with the problem considered when
we pass from the 2D geometry to the quasi-1D one by rolling the plane into a
cylinder.

\subsubsection{Temperature dependence}
\label{s4.3.1}

In the presence of interactions, the temperature $T$ plays a twofold role,
similarly to the frequency. On one hand, it induces an averaging over the
energy window of the width $\sim T$, thus breaking the chiral symmetry and
``switching on'' the localization effects. On the other hand, the interaction
at finite $T$ generates a non-zero dephasing rate $\tau_{\phi}^{-1}(T)$ cutting
off the localization corrections. As we showed in Sec.\ \ref{s4.2}, the level 
spacing $\Delta_\xi(T)$ is $\sim T$, so that the result of the competition of
these two effects depend on the value of $T\tau_\phi(T)$. The theory of
dephasing in the present situation remains to be developed. If the dominant
mechanism of dephasing is the electron-electron interaction, one can expect
that (like in conventional 2D systems with dimensionless conductivity replaced
by unity)  $\tau_\phi^{-1}(T) \sim T$. If this is indeed true, the behavior of
the conductivity at $T<\Gamma$ will be qualitatively analogous to that for the
case of finite frequency, Sec.\ \ref{s4.2}. At high temperatures, $T>\Gamma$,
the $T$-dependence of the conductivity will essentially reproduce its 
$\varepsilon$ dependence for given type of disorder (Born or unitary). 

More realistically, one can think about a situation when the disorder is
predominantly chiral, but the chiral symmetry is slightly
broken,\cite{Altland03, Yashenkin01} e.g., by weak potential disorder on the
energy scale $\Gamma_\chi \ll \Gamma$. Then the above consideration allowing
one to expect the conductivity $\sim e^2/h$ will be applicable in the intermediate
range, $\Gamma_\chi < T < \Gamma$;  at still lower temperatures, $T \ll
\Gamma_\chi$, the chirality-breaking effects will drive the system into the
strong localization regime. 

It is worth noting that the interaction may lead to other effects (in
particular, to open the gap in the spectrum and/or to break the chiral
symmetry, cf. Refs.\ \onlinecite{Foster06,Ludwig06}) not included in our
consideration. These questions also require further study.

\subsubsection{Mesoscopic sample}
\label{s4.3.2}

Let us now consider the situation when all the potential infrared regulators
$\delta,T,\omega$ are much smaller than the level spacing in the sample. In
this case, the sample will be fully phase-coherent (mesoscopic) and its size
will serve as an infrared cut-off. In such a mesoscopic situation one should in
general speak about a conductance, not conductivity. Furthermore, the
properties of the conductance will essentially depend on the sample geometry.
Consider a rectangular sample $L_x \times L_y$, with current flowing along the
$x$ axis. For an approximately square sample, $L_x\sim L_y$, we expect, based
on the above results, an average conductance of order $e^2/h$. Indeed, one can
imagine taking first $\delta$ larger than the level spacing (so that the result
of Sec.\ \ref{s4.1} applies) and then decreasing it until it reaches the level
spacing.) The above statement follows from the continuity. In view of the
mesoscopic character of the sample, we also expect in this case a broad
distribution of the conductance, with a variance of the order of  $(e^2/h)^2$.
Both the average value and the conductance distribution will depend on the
exact value of the aspect ratio $L_x/L_y$ in a non-trivial way and can hardly
be calculated.

For a long sample, $L_x\gg L_y$, the geometry becomes quasi-one-dimensional,
and our results for the universal conductivity $\sim e^2/h$ cease to be
relevant (see also Sec.\ \ref{s4.3.3}).

Finally, let us consider a case of a very broad and short sample, $L_y \gg
L_x$. In this situation, the conductance will be self-averaging, so that one
can again speak about conductivity. Using again the continuity, we conclude
that the conductivity in this situation will have some universal value
$\sigma_L \sim e^2/h$. Whether this value is equal to the above universal
conductivity $4e^2/\pi h$, Eq.\ (\ref{universal}), or the numerical coefficient
is different, requires  further study. Remarkably, the same value Eq.\
(\ref{universal}) has been found\cite{Katsnelson, Tworzydlo06, Ryu} for the
conductance of a {\it clean} graphene sample in the considered geometry $L_y\gg
L_x$.

\subsubsection{Cylindric geometry}
\label{s4.3.3}

Let us take a $C_0$-symmetric strip of a large transverse size $L_y$ and
infinite in the $x$ direction, and roll into a cylinder, preserving the chiral
structure. Let us further assume a small but non-zero level width $\delta$, as
in Sec.\ \ref{s4.1}. If $\delta$ is much larger than the level spacing in the
square $L_y \times L_y$, the system is effectively two-dimensional, and the
consideration of Sec.\ \ref{s4.1} applies. Let us consider the opposite limit.
One can then ask whether  our result concerning the universal conductivity will
be applicable in this quasi-1D geometry. Analyzing the derivation in Sec.\
\ref{s4.1} it is not difficult to see that it breaks down: the momentum $q_y$
is now quantized and its shift therefore not allowed. Let us consider,
however, an Aharonov-Bohm flux $\Phi$ piercing the cylinder, which amounts to
introducing the extra phase $e^{i\Phi/\Phi_0}$ in the periodic boundary
conditions ($\Phi_0$ is the flux quantum). Averaging over $\Phi$, we restore
the applicability of the consideration of Sec.\ \ref{s4.1}, so that
\begin{equation}
 \langle \sigma \rangle_\Phi
  = \frac{4e^2}{\pi h}.
\end{equation}
Therefore, depending on the value of the Aharonov-Bohm flux, the conductivity
can be either larger or smaller than this universal value, which is 
restored after the averaging. A similar strong dependence of conductance of a
clean graphene strip on the boundary conditions was found in Refs.\
\cite{Katsnelson, Tworzydlo06}. Our observation of the Aharonov-Bohm flux
dependence of the conductivity seems also to be related to the known results on
transport properties of disordered wires with chiral symmetry, namely their
dependence on the parity of the number of channels and the staggering in the
hopping matrix elements.\cite{Brouwer}

\section{Conclusions}
\label{s5}

To summarize, we have studied electron transport properties of a disordered
graphene layer. We have shown that the nature of disorder is of crucial
importance for the behavior of the conductivity. Away from the half filling,
the concentration dependence of conductivity is linear (with logarithmic
corrections) for strong scatterers (unitary limit), while it is only
logarithmic in the case of weak scatterers (Gaussian disorder). We have
constructed a ``phase diagram'' showing which of these types of behavior should
be expected for given microscopic parameters of the disorder. We have shown
that the physically important case of charged impurities corresponds to
the Gaussian-unitary ``phase
boundary''. The linear behavior of the conductivity that we have found for the
case of strong scatterers agrees with the experimental
findings,\cite{Novoselov05, Zhang} demonstrating that this kind of disorder is
dominant in experimentally studied structures.

At half filling, the conductivity is of the order of $e^2/h$ if the randomness
preserves one of the chiral symmetries of the clean Hamiltonian; otherwise, the
conductivity is strongly affected by localization effects. For the case of
chiral disorder, the exact value of the conductivity still depends on the
nature of the infrared cut-off, which may depend on the physical setup. We have
analyzed in detail the situation when this cut-off is provided by the level
width $\delta$ or by the frequency $\omega$; in the first case the conductivity
takes a universal value $4 e^2/\pi h$, while in the second case it shows a more
complex behavior. Whether the chiral disorder may indeed dominate in
experimentally relevant structures, explaining the observed value of
conductivity $\sim e^2/h$ remains an open question. From the theoretical point
of view, further research directions extending our results include, in
particular, the mesoscopic transport in a phase-coherent disordered sample and
effects of interaction.

\begin{acknowledgments}
We thank D.I.~Diakonov, I.A.~Gruzberg, A.W.W.~Ludwig, K.S.~Novoselov,
S.V.~Morozov, A.F.~Morpurgo, M.A.~Skvortsov, and A.G.~Yashenkin for valuable
discussions. We are also grateful to A.W.W.~Ludwig for bringing Ref.\
\onlinecite{BernardLeClair} to our attention. The work was supported by the
Center for Functional Nanostructures and the Schwerpunktprogramm
``Quanten-Hall-Systeme'' of the Deutsche Forschungsgemeinschaft. The work of
PMO was supported by the Russian Foundation for Basic Research under grant No.\
04-02-16348 and by the Russian Academy of Sciences under the program ``Quantum
Macrophysics.'' The work of IVG, conducted as a part of the project ``Quantum
Transport in Nanostructures'' made under the EUROHORCS/ESF EURYI Awards scheme,
was supported by funds from the Participating Organizations of EURYI and the EC
Sixth Framework Programme and by the Program ``Leading Russian Scientific
Schools'' under Grant No.\ 2192.2003.2. ADM acknowledges hospitality of the
Kavli Institute for Theoretical Physics at Santa Barbara during the completion
of the manuscript and partial support by the National Science Foundation under
Grant No.\ PHY99-07949.
\end{acknowledgments}

\begin{widetext}

\appendix

\section{One-loop RG equations}
\label{App:RG}

\allowdisplaybreaks[4]

A complete set of one-loop perturbative RG equations can be obtained by
considering the diagrams of Fig.\ \ref{Fig:RG} with all possible disorder
structures from Table \ref{Tab:sym}. An impurity line in those diagrams
represents a sum of all possible types of disorder with the proper amplitude
and corresponding matrices at the vertices. The RG equations for 9 disorder
amplitudes [diagrams (b), (c), and (d) in Fig.\ \ref{Fig:RG}] have the form
\begin{subequations}
\begin{align}
 \frac{d \alpha_0}{d \log L}
  &= 2 \alpha_0 (\alpha_0 + \beta_0 + \gamma_0 + \alpha_\perp + \beta_\perp
     + \gamma_\perp + \alpha_z + \beta_z + \gamma_z) + 2 \alpha_\perp \alpha_z
     + \beta_\perp \beta_z + 2\gamma_\perp \gamma_z, \\
 \frac{d \alpha_\perp}{d \log L}
  &= 2 (2 \alpha_0 \alpha_z + \beta_0 \beta_z + 2 \gamma_0 \gamma_z), \\
 \frac{d \alpha_z}{d \log L}
  &= -2 \alpha_z (\alpha_0 + \beta_0 + \gamma_0 - \alpha_\perp - \beta_\perp
     - \gamma_\perp + \alpha_z + \beta_z + \gamma_z) + 2 \alpha_0 \alpha_\perp
     + \beta_0 \beta_\perp + 2 \gamma_0 \gamma_\perp, \\
 \frac{d \beta_0}{d \log L}
  &= 2 [\beta_0 (\alpha_0 - \gamma_0 + \alpha_\perp + \alpha_z - \gamma_z)
     + \alpha_\perp \beta_z + \alpha_z \beta_\perp + \beta_\perp \gamma_0], \\
 \frac{d \beta_\perp}{d \log L}
  &= 4 (\alpha_0 \beta_z + \alpha_z \beta_0 + \beta_0 \gamma_0
     + \beta_\perp \gamma_\perp + \beta_z \gamma_z), \\
 \frac{d \beta_z}{d \log L}
  &= 2 [-\beta_z (\alpha_0 - \gamma_0 - \alpha_\perp + \alpha_z - \gamma_z)
     + \alpha_0 \beta_\perp + \alpha_\perp \beta_0 + \beta_\perp \gamma_z], \\
 \frac{d \gamma_0}{d \log L}
  &= 2 \gamma_0 (\alpha_0 - \beta_0 + \gamma_0 + \alpha_\perp - \beta_\perp
     + \gamma_\perp + \alpha_z - \beta_z + \gamma_z) + 2 \alpha_\perp \gamma_z
     + 2 \alpha_z \gamma_\perp + \beta_0 \beta_\perp, \\
 \frac{d \gamma_\perp}{d \log L}
  &= 4 \alpha_0 \gamma_z + 4 \alpha_z \gamma_0 + \beta_0^2 + \beta_\perp^2
     + \beta_z^2, \\
 \frac{d \gamma_z}{d \log L}
  &= -2 \gamma_z (\alpha_0 - \alpha_\perp + \alpha_z - \beta_0 + \beta_\perp
     - \beta_z + \gamma_0 - \gamma_\perp + \gamma_z) + 2 \alpha_0 \gamma_\perp
     + 2 \alpha_\perp \gamma_0 + \beta_\perp \beta_z. \\
\intertext{The RG equation for the energy [diagram (a) in Fig.\ \ref{Fig:RG}]
reads}
 \frac{d \varepsilon}{d \log L}
  &= \varepsilon(1 + \alpha_0 + \beta_0 + \gamma_0 + \alpha_\perp + \beta_\perp
     + \gamma_\perp + \alpha_z + \beta_z + \gamma_z).
\end{align}
\label{completeRG}
\end{subequations}
For brevity, in this Appendix we omit tildes which distinguish running
parameters from their initial values in the main text.

In various particular cases, when only some subset of disorder structures is
present, these equations reduce to the corresponding form known in the
literature. The cases of $C_0$-chiral ($\alpha_\perp$, $\beta_\perp$,
$\gamma_\perp$) and $C_z$-chiral ($\alpha_\perp$, $\beta_0$, $\beta_z$,
$\gamma_\perp$) disorder are considered in Sec.\ \ref{s4.1}. If the disorder is
proportional to the $\tau_3$ matrix ($\alpha_\perp$, $\gamma_0$, $\gamma_z$),
the Hamiltonian decouples in two $2 \times 2$ blocks, which have the structure
of the model with random mass ($\gamma_z$), scalar ($\gamma_0$), and vector
($\alpha_\perp$) potential analyzed in Ref.\ \onlinecite{Ludwig}. The RG
equations for the random mass problem were also given in Refs.\
\onlinecite{Dotsenko83,Bocquet}, for the random potential in Ref.
\onlinecite{NersesyanTsvelik}.

If the system possesses a time-reversal invariance ($T_0$), only the couplings
$\alpha_0$, $\beta_\perp$, $\beta_z$, $\gamma_\perp$, and $\gamma_z$ survive,
which is the case considered in Ref.\ \onlinecite{AleinerEfetov}. Taking into
account the difference between our RG scheme and that of Ref.\
\onlinecite{AleinerEfetov} (where the velocity is renormalized whereas the
energy is not), we have checked that RG equations of Ref.\
\onlinecite{AleinerEfetov} are reproduced from the complete set
(\ref{completeRG}) if a number of assumptions concerning the hierarchy of the
disorder couplings ($\alpha_0 \gg \sqrt{\alpha_0 |2\beta_z - \beta_\perp|},
\sqrt{\alpha_0 |2\gamma_z - \gamma_\perp|} \gg \beta_z, \beta_\perp,
\gamma_z, \gamma_\perp \gg |2\beta_z - \beta_\perp|, |2\gamma_z -
\gamma_\perp|$) are made.

\section{Impurity-induced corrections to the DoS in the unitary limit}
\label{App:unitary_rho}

In this Appendix we calculate the density of states in the presence of
infinitely strong impurities (unitary limit) up to the second order in their
concentration $n_\text{imp}$. The contribution of the first order in
$n_\text{imp}$ is determined by the diagram (Fig.\ \ref{Fig:uni-diagrams}a)
containing a single $T$-matrix\cite{footnotePepinLee}
\begin{equation}
 \delta\rho^{(1)}(\varepsilon)
  = -\frac{4}{\pi} \mathop{\mathrm{Im}} 
    \frac{\eta \Delta^2}
         {2[\varepsilon \log(\Delta/|\varepsilon|) + i\pi|\varepsilon|/2]}
    \int \frac{d^2 k}{(2\pi)^2}\, \frac{\varepsilon^2 + v_0^2 k^2}
     {\left[(\varepsilon + i0)^2 - v_0^2 k^2 \right]^2}
  = -\frac{\eta \Delta^2}
          {2\pi v_0^2 |\varepsilon| \log^2(\Delta/|\varepsilon|)}.
 \label{delta_rho_1_unitary}
\end{equation}
This result corresponds to Eq.\ (\ref{dos-unitary-1}). Obviously, the
first-order contribution to the density of states is correctly taken into
account by the self-consistent unitary approximation.

The problem becomes more complicated when one looks for the second-order
contribution. The calculations are greatly simplified in the coordinate
representation and for Matsubara energies. The Green function and the
$T$-matrix have the form
\begin{gather}
 G_0(i\epsilon, \mathbf{r})
  = -\frac{i\epsilon}{2\pi v_0^2}\left[
      K_0\left(\frac{\epsilon r}{v_0}\right)
      + \tau_3 \bm{\sigma} \hat{\mathbf{r}}
        K_1\left(\frac{\epsilon r}{v_0}\right)
    \right], \label{Green-eps-r}\\
 T(i\epsilon)
  = \frac{2\pi v_0^2}{i b \epsilon \log(\Delta/\epsilon)},
 \qquad\qquad   
 b   
  = \begin{cases}
      1, &\text{long-range}; \\
      4, &\text{short-range}.
    \end{cases}
 \label{TMatsubara}
\end{gather}
The $T$-matrix has different values in the limits of long- and short-range
potential disorder. Eq.\ (\ref{Green-eps-r}) for the Green function applies
for not too short distance. One has to cut the real-space integrals at $r \sim
v_0/\Delta$.

We are going to express the density of states in terms of the partition
function per unit area. The contribution to this quantity of the second order
in $\eta$ is given by the diagrams Fig. \ref{Fig:uni-diagrams}b
\begin{equation}
 Z_2(i\epsilon)
  = n_\text{imp}^2 \mathop{\mathrm{Tr}} \sum_{m=1}^\infty
    \frac{T^{2m}}{2m} \int d^2 r [G(r)G(-r)]^m
  = -2 n_\text{imp}^2 \int d^2r \log \left\{
      1 - \left(\frac{T \epsilon}{2\pi v_0^2}\right)^2 \left[
        K_1^2\left(\frac{\epsilon r}{v_0}\right)
        -K_0^2\left(\frac{\epsilon r}{v_0}\right)
      \right]
    \right\}.
\end{equation}

The correction to the density of states can be represented in the form
\begin{equation}
 \delta\rho^{(2)}(\varepsilon)
  = -\frac{1}{\pi} \mathop{\mathrm{Im}} \left.
      \frac{d Z_2}{d(i\epsilon)}
    \right|_{i\epsilon \to \varepsilon + i0}
  = 8 v_0^2 n^2_\mathrm{imp} \mathop{\mathrm{Im}} \left[
      \frac{i}{\epsilon^3} \int \frac{dz\; z\, K_1^2 (z)}
        {b^2 \log^2(\Delta/\epsilon) + K_1^2 (z) - K_0^2(z)}
    \right]_{i\epsilon\to \varepsilon+i0}.
 \label{delta_rho_2_integral}
\end{equation}
For $\mathcal{L} = b \log(\Delta/\epsilon) \gg 1$ we split the integral over
$z$ into two parts and observe that it is dominated by the domain $z >
1/\mathcal{L}$:
\begin{equation}
 \int_0^{\infty} \frac{dz\; z\, K_1^2(z)}{\mathcal{L}^2 + K_1^2(z) - K_0^2(z)}
  \simeq \int_0^{1/\mathcal{L}} dz\; z
    + \frac{1}{\mathcal{L}^2} \int_{1/\mathcal{L}}^{\infty} dz\; z\, K_1^2(z)
  \simeq \frac{\log \mathcal{L}}{\mathcal{L}^2}.
\end{equation}
Substituting this in Eq.\ (\ref{delta_rho_2_integral}) and performing the
analytical continuation, we finally arrive at
\begin{multline}
 \delta\rho_2(\varepsilon)
  = 8 v_0^2 n^2_{\mathrm{imp}} \mathop{\mathrm{Im}} \left[
      \frac{i}{\epsilon^3}
      \frac{\log\log(\Delta/\epsilon)}{b^2 \log^2(\Delta/\epsilon)}
    \right]_{i\epsilon\to \varepsilon+i0}
  = -\frac{8 \pi v_0^2 n^2_{\mathrm{imp}}}{b^2 |\varepsilon|^3}
    \frac{\log\log(\Delta/|\varepsilon|)}{\log^3(\Delta/|\varepsilon|)} 
  = -2\rho(\varepsilon) \alpha_U^2(\varepsilon)
    \log\frac{\Delta}{|\varepsilon|}
    \log\log\frac{\Delta}{|\varepsilon|},
 \label{delta_rho_2_unitary}
\end{multline}
where $\alpha_U(\varepsilon)$ is determined by Eq.\ (\ref{alpha_U}).

\begin{figure}
\includegraphics[width=0.5\columnwidth]{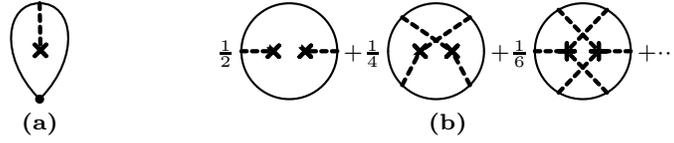}
\caption{Diagrams for (a) the first-order correction to the density of states
and (b) the second-order correction to the partition function.}
\label{Fig:uni-diagrams}
\end{figure}

\end{widetext}


\begin{thebibliography}{99}
\bibitem{footnoteAE}
Since Ref.\ \onlinecite{AleinerEfetov} uses a representation different from
Eq.\ (\ref{4D}), and thus the form of the bare Dirac Hamiltonian different from
Eq.\ (\ref{ham}), the identification of the types of disorder requires the
corresponding unitary transformation.
\bibitem{SCBARefs}
In the context of Dirac fermions in disordered superconductors the SCBA scheme
was developed in Ref.\ \onlinecite{Gorkov85} (density of state) and Ref.\
\onlinecite{Lee93} (transport).
\bibitem{footnote1}
In fact, a careful analysis of the SCBA equation yields $\alpha\Gamma_0 =
\Gamma_0 / \log(\Delta/\Gamma_0)$ for the position of the crossover between the
high- and the low-energy asymptotics in Eq.\ (\ref{SigmaSCBA}).  For
simplicity, we omit such logarithmic prefactors in the expressions for the
boundary between the high- and the low-energy regimes in Eq.\ (\ref{SigmaSCBA})
and below.
\bibitem{footnote-current}
It is worth mentioning that the current operator, $e v_0 \bar\psi \tau_3
\bm{\sigma} \psi$, is not renormalized within the considered ultraviolet
(ballistic) RG, apart from its engineering dimension. Indeed, the
retarded-retarded current operator $e v_0 \bar\psi^R \tau_3 \bm{\sigma}
\psi^R$, is related to the kinetic term $i v_0 \bar\psi \tau_3 \bm{\sigma}
\nabla \psi$ by virtue of gauge invariance and thus does not acquire any
anomalous scaling. Further, since the ultraviolet RG works identically for
RR and RA operators, this statement applies to the latter as well.
\bibitem{footnote2}
Another feature of such a crossover --- also not observed experimentally ---
would be the asymmetric energy dependence of the density of
states\cite{Pogorelov06} and the conductivity in the realistic case of positive
impurity potential. In the Born limit, the chemical potential is shifted by
$\delta\mu=n_\text{imp}U \sim \Delta(\alpha\eta)^{1/2}$, whereas in the
unitary case this shift vanishes. Also, the density of states would exhibit a
resonant structure related to emergence of bound states.\cite{Pogorelov06,
Wehling, BalatskyRMP}
\bibitem{footnote3}
In Refs.\ \onlinecite{Gade93, Guruswamy} the function $f(\varepsilon)$ was
found to be of the form $f(\varepsilon)\sim\exp\{-c|\log\varepsilon|^x\}$ with
$x=1/2$; later works\cite{Motrunich,MudryDOS} corrected the value of $x$ to
$x=2/3$.
\bibitem{footnote-polarization}
The second term in the denominator of Eq.\ (\ref{coulomb-screened})
(polarization function) is written in the Thomas-Fermi approximation valid for
$q \lesssim \varepsilon/v_0$; for larger momenta it will be suppressed by a
factor $\varepsilon/q v_0 \ll 1$. This is, however, not important, since for
such $q$ the first term (bare interaction) dominates anyway.
\bibitem{footnote-screen}
The Thomas-Fermi approach implies the locality of the nonlinear response
function $\Pi_3(\mathbf{r},\mathbf{r}',\mathbf{r}'') = (\partial^2n/\partial
\mu^2) \delta(\mathbf{r} - \mathbf{r}') \delta(\mathbf{r} - \mathbf{r}'')$ [in
the momentum representation $\tilde\Pi_3(\mathbf{q}, \mathbf{q}', \mathbf{q}'')
= \partial^2n/\partial \mu^2 \delta(\mathbf{q} + \mathbf{q}' + \mathbf{q}'')$]
represented by a three-leg fermionic loop. This form is valid only on spatial
scales much larger than the Fermi wavelength. In Ref.\
\onlinecite{Katsnelson-screen}, the Thomas-Fermi approximation has been applied
to the scales shorter than the Fermi wavelength ($a < \mathbf{r'}, \mathbf{r''}
< \lambda_\mu$), i.e. for $\mu \ll q v_0 \ll \Delta$. On such scales the
nonlinear response function $\Pi_3$ is nonlocal in space (in momentum
representation it is suppressed by $1/q^2$) which eliminates all the
ultraviolet logarithmic singularities.
\end{thebibliography}
\end{document}